\pdfoutput=1

\documentclass[11pt,twoside,a4paper,cmspaper,final,collab]{cms-tdr}
\def\svnVersion{347882}\def\cmsCernNoTag{CERN-PH-EP/2015-170}\def\cmsCernDate{\today}\def\cmsMessage{Published in Physical Review D as \href{http://dx.doi.org/10.1103/PhysRevD.93.112009}{\doi{10.1103/PhysRevD.93.112009.}}}

\begin{document}\cmsNoteHeader{B2G-13-006}

\hyphenation{had-ron-i-za-tion}
\hyphenation{cal-or-i-me-ter}
\hyphenation{de-vices}

\RCS$Revision: 347749 $
\RCS$HeadURL: svn+ssh://svn.cern.ch/reps/tdr2/papers/B2G-13-006/trunk/B2G-13-006.tex $
\RCS$Id: B2G-13-006.tex 347749 2016-06-17 11:55:02Z jpilot $
\newlength\cmsFigWidth
\ifthenelse{\boolean{cms@external}}{\setlength\cmsFigWidth{0.98\columnwidth}}{\setlength\cmsFigWidth{0.6\textwidth}}
\ifthenelse{\boolean{cms@external}}{\providecommand{\cmsLeft}{top}}{\providecommand{\cmsLeft}{left}}
\ifthenelse{\boolean{cms@external}}{\providecommand{\cmsRight}{bottom}}{\providecommand{\cmsRight}{right}}
\ifthenelse{\boolean{cms@external}}{\providecommand{\CLns}{C.L\xspace}}{\providecommand{\CLns}{CL\xspace}}
\ifthenelse{\boolean{cms@external}}{\providecommand{\CL}{C.L.\xspace}}{\providecommand{\CL}{CL\xspace}}
\newlength\cmsTabSkip\setlength\cmsTabSkip{1.5ex}
\newcommand{\bprime}{\PB\xspace}
\newcommand{\baprime}{\PaB\xspace}
\newcommand{\bprimetotW}{\ensuremath{\bprime\to \cPqt \PW}\xspace}
\newcommand{\bprimetobZ}{\ensuremath{\bprime\to \cPqb \Z}\xspace}
\newcommand{\bprimetobH}{\ensuremath{\bprime\to \cPqb \PH}\xspace}
\newcommand{\boxCheck}{$\text{\rlap{$\checkmark$}}\square$}
\newcommand{\ttjets}{\ttbar{}+jets\xspace}
\newcommand{\mellellb}{\ensuremath{M(\ell\ell\cPqb)}\xspace}
\newcommand{\Zelel}{\ensuremath{\Z \to \Pep\Pem}\xspace}
\newcommand{\Zmumu}{\ensuremath{\Z \to \Pgmp\Pgmm}\xspace}
\newcommand{\bjet}{\cPqb\text{ jet}\xspace}
\newcommand{\bjets}{\cPqb\text{ jets}\xspace}
\newcommand{\Hbb}{\ensuremath{\PH\to\cPqb\cPaqb}\xspace}
\newcommand{\mpruned}{\ensuremath{M_\text{pruned}}\xspace}
\newcommand{\TWEXPLIMIT}{890\xspace}
\newcommand{\BHEXPLIMIT}{810\xspace}
\newcommand{\BZEXPLIMIT}{740\xspace}
\newcommand{\TWOBSLIMIT}{880\xspace}
\newcommand{\BHOBSLIMIT}{900\xspace}
\newcommand{\BZOBSLIMIT}{750\xspace}
\providecommand{\ST}{\ensuremath{S_{\mathrm{T}}}\xspace}
\providecommand{\tauh}{\ensuremath{\Pgt_\mathrm{h}}\xspace}
\newcolumntype{x}{D{,}{\pm}{-1}} 

\cmsNoteHeader{B2G-13-006}
\title{Search for pair-produced vectorlike \bprime quarks in proton-proton collisions at \texorpdfstring{$\sqrt{s} = 8$\TeV}{sqrt(s) = 8 TeV}}

\date{\today}

\abstract{A search for the production of a heavy \bprime quark, having electric charge ${-}1/3$ and vector couplings to \PW, \Z, and \PH bosons, is carried out using proton-proton collision data recorded at the CERN LHC by the CMS experiment, corresponding to an integrated luminosity of 19.7\fbinv. The \bprime quark is assumed to be pair produced and to decay in one of three ways: to $\PQt\PW$, $\PQb\Z$, or $\PQb\PH$. The search is carried out in final states with one, two, and more than two charged leptons, as well as in fully hadronic final states. Each of the channels in the exclusive final-state topologies is designed to be sensitive to specific combinations of the \bprime quark-antiquark pair decays. The observed event yields are found to be consistent with the standard model expectations in all the final states studied. A statistical combination of these results is performed, and upper limits are set on the cross section of the strongly produced \bprime quark-antiquark pairs as a function of the \bprime{} quark mass. Lower limits on the \bprime quark mass between 740 and \BHOBSLIMIT\GeV are set at a 95\% confidence level, depending on the values of the branching fractions of the \bprime quark to $\PQt\PW$, $\PQb\Z$, and $\PQb\PH$. Overall, these limits are the most stringent to date.

}

\hypersetup{%
pdfauthor={CMS Collaboration},%
pdftitle={Search for pair-produced vectorlike B quarks in proton-proton collisions at sqrt(s) = 8 TeV},%
pdfsubject={CMS},%
pdfkeywords={CMS, physics, B2G, beyond two generations, vector-like quark, bprime}}

\maketitle

\section{Introduction}\label{sec:introduction}

The discovery of a Higgs boson \cite{Chatrchyan:2013lba,Aad:2012tfa,Chatrchyan:2012ufa} has intensified the search for physics beyond the standard model (SM). Various extensions of the SM predict the existence of new heavy quarks, which arise quite naturally in grand unification schemes~\cite{Bagger:1984gn} and in composite Higgs~\cite{Agashe:2004rs,Contino:2003ve}, little Higgs~\cite{Berger:2012ec,Han:2005ru,Perelstein:2003wd,ArkaniHamed:2002qx}, and top quark condensate~\cite{Dobrescu:1997nm} models. The couplings to the SM gauge bosons of the left- and right-handed components of these quarks are symmetric, so they are called vectorlike~\cite{Aguilar-Saavedra:2013qpa}. Vectorlike quarks may be singlets, doublets, or triplets under the electroweak $SU(2)\times U(1)$ transformation~\cite{Fajfer:2013wca}. They have bare mass terms that are invariant under the electroweak gauge transformation~\cite{Frampton:1999xi}. Moreover, their couplings to the scalar sector are independent of mass. Thus, the existence of vectorlike quarks is not ruled out by the recent discovery of a Higgs boson, in contrast to additional quarks in more conventional fourth-generation models~\cite{Djouadi2012310}.

In several beyond the standard model scenarios, vectorlike quarks are considered partners of the top and bottom quarks~\cite{Gopalakrishna:2011ef}. Both the charged-current~\cite{AguilarSaavedra:2009es} and the flavor changing neutral current (FCNC) \cite{Okada:2012gy,FCNC_QtoqZ} decay processes are allowed. The ratio of the predicted rates depends on the model: in some models the FCNC process dominates \cite{PRD2010_81_035004}; in others the two modes are comparable in rate.

\begin{figure}[h!tb]
\centering
\includegraphics[width=0.4\textwidth]{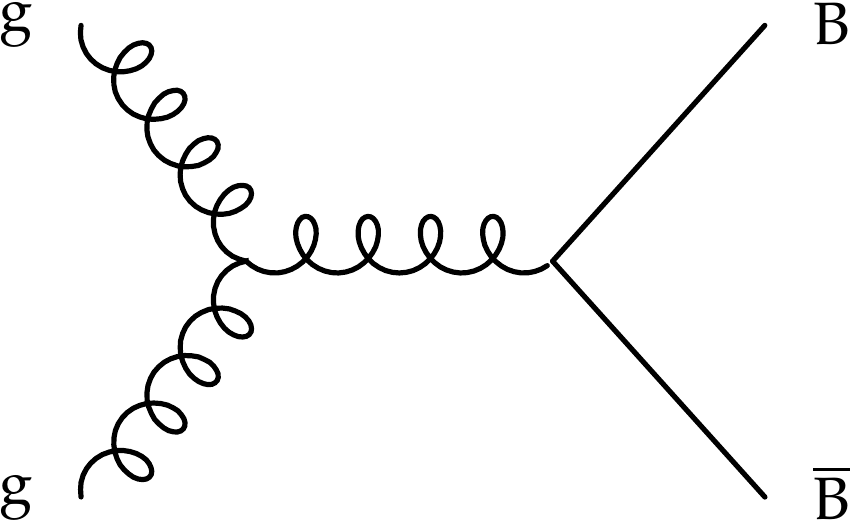} \\
\includegraphics[width=0.2\textwidth]{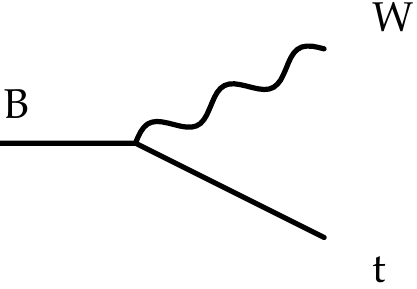}
\includegraphics[width=0.2\textwidth]{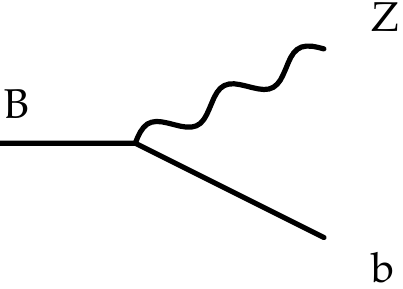}
\includegraphics[width=0.2\textwidth]{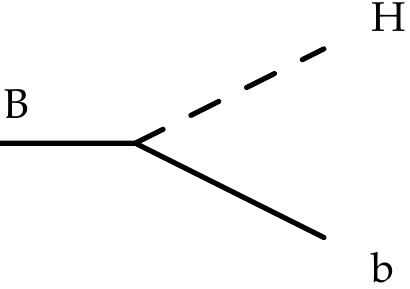}
\caption{Feynman diagrams for the dominant \bprime quark pair-production process (top) and for the \bprime quark decay modes (bottom).}
\label{diagrams}

\end{figure}

This paper describes the search for a vectorlike \bprime{} quark of electric charge ${-}1/3$, using data recorded by the CMS experiment from proton-proton collisions at a center-of-mass energy of 8\TeV at the CERN LHC in 2012. It is assumed that \bprime quark-antiquark pairs are produced strongly for \bprime quark masses within the range of this search, which extends to 1\TeV. The \bprime{} quark may decay either via the charged-current process \bprimetotW or via the FCNC processes \bprimetobZ and \bprimetobH.  Feynman diagrams for the \bprime quark pair production and decay processes are shown in Fig.~\ref{diagrams}.   Searches are performed in several different final states, including those containing single leptons, lepton pairs (dileptons) of opposite or identical charge, three or more leptons, or consisting entirely of hadronic activity without any identified leptons. We search for an excess of events over the backgrounds in mutually exclusive final states and set limits on the pair-production cross section for all values of the \bprime{} quark branching fractions with the constraint $\mathcal{B}(\bprimetotW) + \mathcal{B}(\bprimetobZ) + \mathcal{B}(\bprimetobH)  = 1$.

Experimental searches for a vectorlike \bprime{} quark have previously been reported by experiments at the Fermilab Tevatron and the CERN LHC. A 95\% \CL lower limit on the mass of the \bprime{} quark was set at 268\GeV by the CDF Collaboration~\cite{Aaltonen:2007je}. Recently the ATLAS Collaboration set lower limits on the \bprime{} quark mass ranging from 575 to 813\GeV, for different combinations of the \bprime{} quark branching fractions~\cite{Aad:2015kqa}.  The present analysis improves upon these existing results, setting the most stringent limits to date on the mass of the vectorlike \bprime{} quark.

The paper is divided into several sections.  Section~\ref{sec:cmsdetector} gives an overview of the CMS detector.  Section~\ref{sec:sigmodels} describes the details of the simulations used for signal and background processes.  Section~\ref{sec:evtreco} describes the reconstruction of physics objects and the event selections specific to each individual channel considered in this analysis.  Section~\ref{sec:bkgestimation} describes background estimation techniques for each of the channels, as well as the specific methods used to discriminate the \bprime{} quark signal from the background, while Sec.~\ref{sec:systematics} describes the systematic uncertainties evaluated for each channel and their treatment in combination.  Finally, Sec.~\ref{sec:combination} provides further details on the combination of analysis channels and Sec.~\ref{sec:results} presents the results obtained from this analysis.  A summary is presented in Sec.~\ref{sec:conclusion}.

\section{CMS detector}\label{sec:cmsdetector}

The central feature of the CMS apparatus is a superconducting solenoid of 6\unit{m} internal diameter, providing a magnetic field of 3.8\unit{T}. Within the solenoid volume are a silicon pixel and strip tracker, a lead tungstate crystal electromagnetic calorimeter (ECAL), and a brass and scintillator hadron calorimeter, each composed of a barrel and two end cap sections. Muons are measured in gas-ionization detectors embedded in the steel flux-return yoke outside the solenoid. Extensive forward calorimetry complements the coverage provided by the barrel and end cap detectors. The first level of the CMS trigger system, composed of custom hardware processors, uses information from the calorimeters and muon detectors to select the most interesting events in a fixed time interval of less than 4\mus. The high-level trigger processor farm further decreases the event rate from around 100\unit{kHz} (the maximum allowed output from the first level) to around 400\unit{Hz}, before data storage.  A more detailed description of the CMS detector, together with a definition of the coordinate system used and the relevant kinematic variables, can be found in Ref.~\cite{Chatrchyan:2008zzk}.

\section{Signal and background simulation}\label{sec:sigmodels}

The following section details the simulation methods used to generate events for modeling the signal and background processes.   One of the main backgrounds in many of the channels is SM \ttbar production.  This process is simulated with the \MADGRAPH v5.1.1 event generator \cite{Maltoni:2002qb}, using the CTEQ6L1 parton distribution function (PDF) \cite{Pumplin:2002vw}.  Events are interfaced with \PYTHIA v6 \cite{Sjostrand:2006za} for shower modeling and hadronization.  These simulation methods are used for the \PW+jets and \Z{}+jets samples, in addition to SM \ttbar production.  For \PW+jets and \Z{}+jets events, up to four additional partons are allowed at the matrix element level during generation.

Diboson processes $\PW\PW$, $\PW\Z$, and $\Z\Z$ are generated with \PYTHIA 6.424, and single top quark processes ($\PQt\PW$, $s$-channel, and $t$-channel) are generated using \POWHEG 1.0 \cite{Nason:2004rx, Frixione:2007vw, Alioli:2010xd, Alioli:2009je} and interfaced with \PYTHIA for shower modeling and hadronization.  Both the diboson and single top processes are generated with the CTEQ6M PDF set.  The rare processes $\ttbar\PW$, $\ttbar\Z$, and $\cPqt\cPaqb\Z$ are simulated with \MADGRAPH v5.

Normalizations for the background processes are initially set according to theoretical predictions, and are allowed to vary within the corresponding uncertainties during cross section limit extraction. For \PW+jets and \Z{}+jets processes, we use the calculations found in Refs.~\cite{fewz1, fewz2, fewz3}.  For \ttbar and single top samples, we normalize using cross sections calculated in Refs.~\cite{ttbar_xsec_new} and \cite{singletop_xsec}, respectively. Finally, for diboson and rare processes we use cross sections computed in Refs.~\cite{diboson_xsec} and \cite{ttWxsec,ttZxsec}, respectively.

To model the kinematic properties of the $\Pp\Pp \to \bprime\baprime$ signal process, we use samples of simulated events produced with the \MADGRAPH v5 generator and CTEQ6L1 PDF set, allowing for up to two additional partons in the final state of the hard scatter matrix element.  The generated events are then interfaced with \PYTHIA v6  for parton shower modeling and hadronization.

Samples are generated for \bprime quark masses between 500 and 1000\GeV, in steps of 50\GeV, for each of the six distinct combinations of decay products: $\PQt\PW\PQt\PW$, $\PQt\PW\PQb\Z$, $\PQt\PW\PQb\PH$, $\PQb\Z\PQb\Z$, $\PQb\PH\PQb\Z$, and $\PQb\PH\PQb\PH$.  The standard model final states identical to those listed here are not considered, as the rates are negligible relative to the other background processes.  By reweighting events from these different samples, an arbitrary combination of branching fractions to $\PQt\PW$, $\PQb\Z$, and $\PQb\PH$ can be probed.  To normalize the simulated samples to expected event yields, we use cross sections computed to next-to-next-to-leading order (NNLO) using both \textsc{hathor} \cite{Aliev:2010zk} and \textsc{Top++}2.0 \cite{Cacciari:2011hy}.  The numerical values used for the \bprime quark pair-production cross sections as a function of mass, generated with \textsc{Top++}2.0, are listed in Table \ref{tab:NNLObpXS}.

\begin{table}[h!tb]
\topcaption{Production cross sections for $\Pp\Pp\to \bprime\baprime$, used to normalize simulated signal samples to expected event yields.  The cross sections are computed to NNLO with Top++2.0 \cite{Cacciari:2011hy}.}
\label{tab:NNLObpXS}
\centering
\begin{scotch}{cD{.}{.}{1.5}}
\multicolumn{1}{c}{$M$(\bprime) [\GeVns{}]} &  \multicolumn{1}{c}{Cross section $\sigma$ [pb]}    \\ \hline
450 &  1.153  \\
500 &  0.590  \\
550 &  0.315  \\
600 &  0.174 \\
650 &  0.0999 \\
700 &  0.0585 \\
750 &  0.0350 \\
800 &  0.0213 \\
850 &  0.0132\\
900 &  0.00828\\
950 &  0.00525\\
1000  & 0.00336 \\
\end{scotch}

\end{table}

Finally, to reproduce the LHC running conditions, simulated events are reweighted to match the observed distribution of the number of reconstructed primary vertices per bunch crossing in data.
\section{Event reconstruction}\label{sec:evtreco}

Events from the LHC collision data or from simulation are reconstructed using the particle flow (PF) algorithm~\cite{CMS-PAS-PFT-09-001, CMS-PAS-PFT-10-001}, which collects information from all subdetectors to reconstruct all detected particles in an event. Events are required to have at least one reconstructed vertex. The interaction vertex with the largest sum of the transverse momentum squared $\pt^2$ of associated tracks is considered the primary interaction vertex. Charged particles originating from other vertices due to additional inelastic proton-proton collisions within the same bunch crossing (pileup) are rejected.

Electron candidates are reconstructed from clusters of energy deposited in the ECAL matched to charged particle trajectories identified in the tracker~\cite{CMS:2013hoa}. Electrons and muons with $\pt$ above 30\GeV and pseudorapidity $\abs{\eta} < 2.4$ are accepted, excluding electrons with $1.44 < \abs{\eta} < 1.57$, in the transition region between the ECAL barrel and end cap. The muon candidates are reconstructed using information from the tracker and the muon spectrometer. Muon candidates are required to have only a small amount of energy deposited in the calorimeters. Further quality requirements are imposed on the muon tracks and the fit to the matched segments in the muon detectors~\cite{CMS-PAPER-MUO-10-004}. Only tracks originating from the primary interaction vertex are considered.

Electron and muon candidates are reconstructed and identified based on quality selection requirements, while hadronically decaying tau leptons (\tauh) are identified using the Hadron Plus Strip (HPS)~\cite{2012JInstHPSTaus} algorithm, which relies on hadrons and photons to construct the various tau lepton hadronic decay modes. The HPS PF tau leptons are required to have $\pt > 20$\GeV and $\abs{\eta} < 2.3$. Additionally, we require \tauh to be separated by $\Delta R = \sqrt{\smash[b]{(\Delta\eta)^{2} + (\Delta\phi)^{2}}} > 0.1$ from electron and muon candidates, where $\phi$ is the azimuthal angle.

The isolation of the lepton candidates (including electrons, muons, and decays of tau leptons to electrons or muons) is measured by the activity in a cone of aperture $\Delta R$ around the lepton direction at the primary vertex. The \pt of charged particles originating at the primary vertex and the \pt of the neutral particles and photons are summed in this cone (excluding the lepton candidate itself) to obtain the isolation variable. Contributions to the neutral hadron and photon energy components due to pileup interactions are subtracted. For the electron isolation, this contribution is determined using the jet area technique~\cite{Cacciari:2007fd}, which computes the transverse energy density of neutral particles using the median of the neutral energy distribution in a sample of jets with $\pt > 3$\GeV.  In the case of the muons, the pileup energy density from neutral particles is estimated to be half of that from charged hadrons, based on measurements performed in jets~\cite{CMS-PAS-PFT-10-002}. The difference between the isolation algorithms arises because electrons and muons are reconstructed using different techniques. Electrons, with large energy deposition in the calorimeters, behave similarly to jets in this respect, while the reconstruction of muons relies more heavily on tracking information. The isolation value, defined as the energy reconstructed in a cone of $\Delta R =0.3$ (0.4) around an electron (muon) candidate, is required to be less than  0.15 (0.12) times the electron (muon) \pt for the lepton to be considered isolated. The lepton identification and isolation conditions remove most of the nonprompt lepton backgrounds.

Particles are clustered to form hadronic jets using the anti-\kt algorithm~\cite{antikT} with a distance parameter of 0.5. Throughout this paper such clusters are referred to as AK5 jets. The AK5 jets with \pt $> 30$\GeV and $\abs{\eta} < 2.4$ are selected, with further requirements that the jet has at least two associated tracks and that at least 1\% of the jet energy fraction is measured in the calorimeters, to remove poorly reconstructed jets. Jet energy corrections are applied; these are derived from simulation and are matched to measurements in data~\cite{Chatrchyan:2011ds}.

Jets arising from the hadronization of \PQb quarks (\PQb jets) are identified using the combined secondary vertex (CSV) \PQb-tagging algorithm~\cite{Chatrchyan:2012btv}, which uses information from tracks and secondary vertices associated with jets to compute a likelihood-based discriminator to distinguish between jets from \PQb quarks and those from charm or light quarks and gluons. The \PQb-tagging discriminator returns a value between 0 and 1, with higher values indicating a higher probability of the jet to originate from a bottom quark. A discriminator threshold is chosen which gives a \PQb-tagging efficiency of about 70\%, with a mistagging rate of about 1.5\% for jets originating from light-flavor quarks or gluons with \pt in the range of 80--120\GeV. The \PQb-tagging efficiency is measured in data and simulation, and corrections are applied to simulated events to account for any differences, as a function of \pt and $\eta$~\cite{CMS:2013vea}. The missing transverse momentum vector is defined as the projection on the plane perpendicular to the beams of the negative vector sum of the momenta of all reconstructed particles in an event. Its magnitude is referred to as $\ETslash$.  The quantity \ST  is defined as the scalar sum of the \pt  of the jets, lepton \pt, and $\ETslash$ in the event.

At very high Lorentz boost, the products of hadronically decaying bosons may be merged into a single reconstructed jet. In this regime, the \PW, \Z, or Higgs bosons are identified as jets clustered with the Cambridge-Aachen algorithm \cite{CACluster1, CACluster2} using a larger distance parameter of 0.8~\cite{CA8}.  In this paper, they are referred to as CA8 jets.  For bosons with \pt above approximately 200\GeV, decay products are expected to be clustered into a single CA8 jet.  Each CA8 jet can be decomposed into constituent subjets using a jet pruning algorithm \cite{Ellis:2009me} to resolve those decay products.  The pruning algorithm removes soft and wide-angle components of the jet during a reclustering, and the last iteration of the clustering process is reversed to identify two subjet candidates within each pruned jet.  Jet properties such as jet mass, $N$-subjettiness \cite{Thaler:2010tr} (used to determine the consistency of a jet with $N$ hypothesized subjets), and the mass drop, defined as the ratio of the most massive subjet to the mass of the pruned jet, are used to identify these bosons.

The trigger selection for each channel entering the combination can be different, depending on the final state of interest.  For the single-lepton channel, two trigger selections are utilized: either a single electron with $\pt > 27$\GeV or a single muon with $\pt > 40$\GeV.  For both of the lepton pair channels, as well as the multilepton channel, three trigger algorithms are used for final states including two electrons, two muons, or one electron and one muon.  In each of these dilepton trigger algorithms, events are selected if the highest-$\pt$ lepton has $\pt > 17$\GeV and the second-highest $\pt$ lepton has $\pt > 8$\GeV.  No charge requirement is applied in the trigger selection, allowing these trigger algorithms to be used in all three channels with two or more leptons.  Finally, the all-hadronic channel uses a trigger algorithm requiring the total scalar $\pt$ sum of reconstructed jets (with $\pt > 30$\GeV and $\abs{\eta} < 3.0$) in the detector to be greater than 750\GeV.  The offline requirements for each channel of the analysis are designed to be fully efficient given these trigger requirements.   Differences in the trigger selections used between analysis channels lead to small differences in the total amount of integrated luminosity utilized in each channel.

The details of the event selections for each individual analysis channel are given in the following subsections.  Table \ref{channels} summarizes these channels in terms of their defining characteristics: the number of selected leptons, the discriminating variable used for limit setting, as well as the decay mode of the \bprime quark for which the channel is most sensitive.

\begin{table*}[htb]
\topcaption{A summary of analysis channels entering the combination, along with the number of selected leptons, the variable used for signal discrimination, and the \bprime quark decay mode providing the best sensitivity for the channel.}
\centering
\begin{scotch}{lccc}
 & Number of leptons & Discriminating variable & Best decay mode \\
\hline
Lepton+jets & 1 & \ST  & $\PQt\PW$ \\
Same-sign dilepton & 2 & \ST  & $\PQt\PW$ \\
Opposite-sign dilepton & 2 & $M(\ell\ell\,\cPqb)$ & $\PQb\Z$ \\
Multilepton & ${\geq}3$ & \ST  & $\PQt\PW$, $\PQb\Z$ \\
All-hadronic & 0 & \HT & $\PQb\PH$ \\
\end{scotch}
\label{channels}
\end{table*}

\subsection{Lepton+jets channel}\label{sec:evtreco_ljets}

Charged leptons from the decays of $\PW$ and $\Z$ bosons are selected using the criteria described in Sec.~\ref{sec:evtreco}, and are required to be isolated from jets. The lepton trajectories are also required to have a transverse impact parameter of less than 0.02\unit{cm} and a longitudinal impact parameter of less than 1\unit{cm} in magnitude, relative to the primary vertex. The final selection requires events to have exactly one isolated lepton and at least four jets with $\pt>200$, 60, 40, 30\GeV, respectively, of which at least one is a bottom jet. The minimum number of jets and the jet \pt requirements are selected to enhance sensitivity to the $\bprime\baprime$ signal with $\bprime\to\PQt\PW$ decays. To further suppress the SM backgrounds, we use the centrality, $C$, defined as the scalar sum of the \pt of the jets divided by the scalar sum of the jet energies, requiring $C>0.4$.   We require events to have $\ETslash > 20$\GeV.  Corrections due to differing trigger, lepton, and \PQb jet identification efficiencies in data and simulation are applied to simulated events.

Events are divided into categories containing 0, 1, or $\geq$2 tagged hadronically decaying \PW, \Z, or Higgs bosons using the CA8 jets. The identification criteria for these signatures require the CA8 jet to have \pt greater than 200\GeV and to be matched to an AK5 jet.  The AK5 jets matched to CA8 jets are then excluded from \PQb-tagging requirements.  The two subjets identified with the pruning algorithm~\cite{Ellis:2009me} are required to have an invariant mass between 50 and 150\GeV, to be consistent with a \PW, \Z, or Higgs boson.  To further reduce SM backgrounds, the mass drop is required to be less than 0.4.  The efficiency of this heavy-boson tagging algorithm is approximately 50\% \cite{JME13006}, and correction factors are applied to compensate for efficiency differences between data and simulation.  To discriminate the $\bprime$ quark signal from the expected backgrounds, the \ST  distribution is used.

\subsection{Same-sign lepton pair channel}\label{sec:evtreco_ss2l}

Events enter the same-sign (SS) dilepton channel if they contain two leptons ($\Pe\Pe$, $\mu\mu$, and $\Pe\mu$) having the same electric charge.  Events containing an additional reconstructed electron, muon, or tau lepton candidate are removed from the final selection.  This channel is optimized for $\bprime\to\PQt\PW$ decays, but maintains some sensitivity for $\PQb\Z$ and $\PQb\PH$ decays.  In events with a top quark and a \PW{} boson, high hadronic activity is expected in addition to the lepton pair, and therefore events are included in the signal region only if they contain four or more jets in addition to the lepton pair.

In this channel the \bprime quarks are not fully reconstructed. Instead, to discriminate signal from background, the \ST  distribution is used.  Events with $\ETslash > 30$\GeV are categorized into five \ST  bins (0.2--0.4, 0.4--0.6, 0.6--0.8, 0.8--1.2, and $\ST > 1.2$\TeV) for each of the dilepton channels.

\subsection{Opposite-sign lepton pair channel}\label{sec:evtreco_os2l}

\begin{figure}
\centering
\includegraphics[width=\cmsFigWidth]{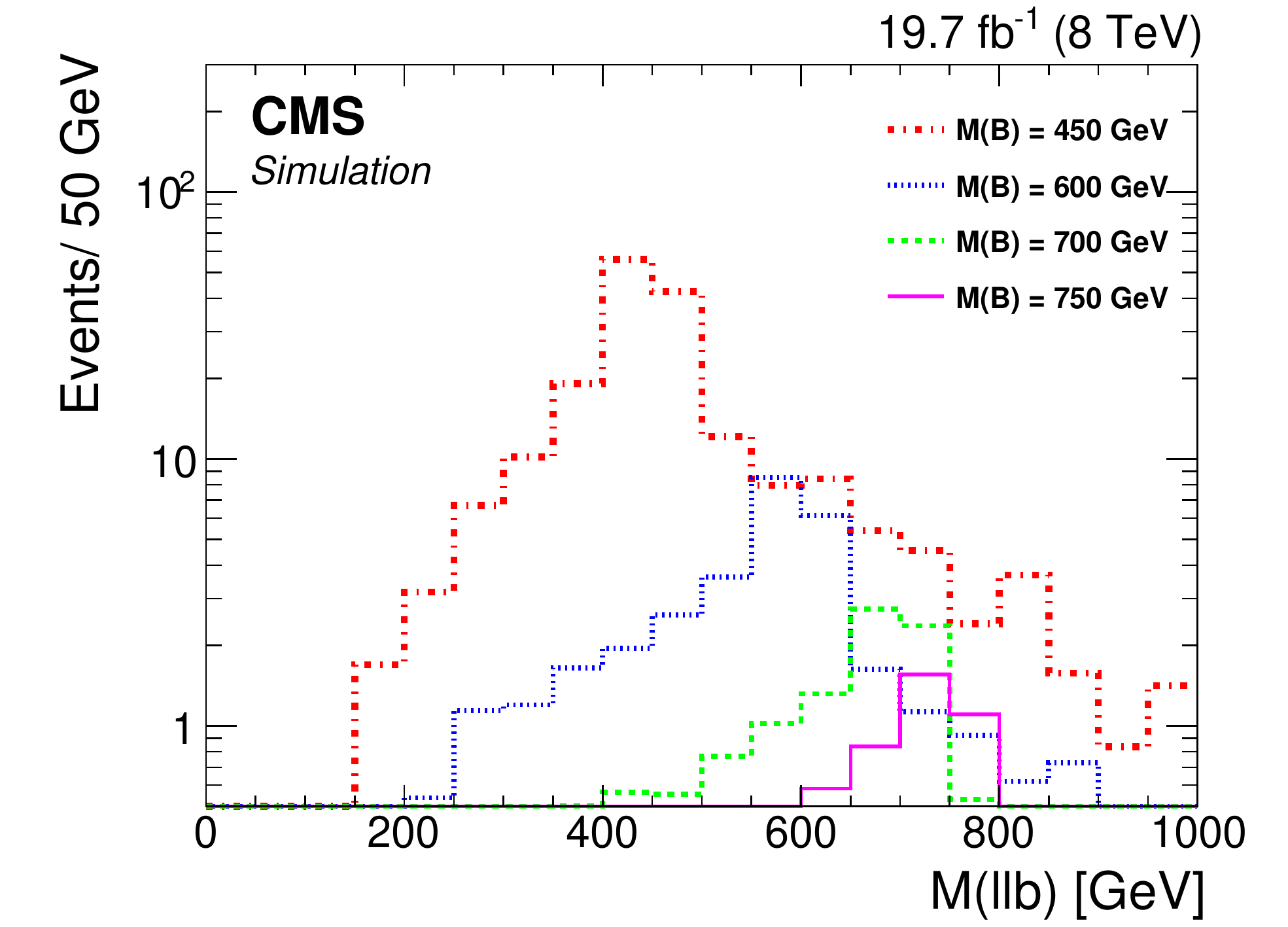}
\caption{The reconstructed mass of the \bprime quark candidate in the opposite-sign lepton pair channel, using the invariant mass of the dilepton and \bjet in simulated events.}
\label{fig:MbZ_MCReco}

\end{figure}

In this channel, optimized for the \bprime{} decaying to a \Z boson and \PQb quark, the \Z boson candidates are reconstructed from pairs of electrons or muons having opposite electric charge, with the identification and isolation criteria previously described. The two highest \pt leptons of the same flavor but opposite charge are used. The pairwise invariant mass of these two objects, $M(\ell\ell)$, where $\ell$ represents an electron or a muon, is required to be in the range of 60--120\GeV, consistent with lepton pairs originating from a \Z boson decay. Furthermore, the $\Z\to\ell^+\ell^-$ candidates are required to have  $\pt(\ell\ell) > 150\GeV$. Events are further required to have at least one \PQb jet  with  $\pt > 80$\GeV. The requirements are optimized to select \Z bosons and \PQb jets originating from the decay of a heavy \bprime{} quark ($>$500\GeV), where the decay products are expected to have large \pt. The kinematic properties of one \bprime{} quark are reconstructed from the \Z boson and the highest-\pt \cPqb{} jet, with $M(\ell\ell\,\PQb)$ used to discriminate the \bprime{} quark signal. The invariant mass distributions of the \bprime quark candidates for different $M(\bprime)$ are shown in Fig.~\ref{fig:MbZ_MCReco}. A signal peak can be identified over a continuous falling background. This reconstruction strategy allows the other \bprime{} quark from the $\bprime\baprime$ quark-antiquark pair to decay into $\cPqb\Z$, $\cPqb\PH$, or $\cPqt\PW$.

\subsection{Multilepton channel}\label{sec:evtreco_multilepton}
Events in this channel must have at least three leptons, consisting of electrons, muons, or tau leptons decaying into fully hadronic states ($\tauh$).  The highest $\pt$ (leading) electron or muon is required to have $\pt > 20$\GeV and the subleading leptons are required to have $\pt > 10$\GeV.  The \pt selection criteria are chosen such that the triggers are fully efficient on these events.

We sort multilepton events into exclusive categories based on the number of leptons, lepton flavor, and relative charges, as well as the kinematic quantity \ST. First, we separate events containing hadronically decaying tau leptons, as their reconstruction is less efficient and this results in lower-purity categories.

Next, we classify each event in terms of the maximum number of opposite-sign and same-flavor (OSSF) lepton pairs that can be made by using each lepton only once. For example, both $\PGmp \PGmm \PGmm$ and  $\PGmp \PGmm \Pem$ contain only one pair of OSSF leptons and are denoted OSSF1, $\PGmp\PGmp\Pem$ contains no such OSSF pair and is denoted OSSF0, while $\PGmp \PGmm \Pep\Pem$ contains two such pairs and is denoted OSSF2. Thus orthogonal categories of events are defined that contain 0, 1, or 2 OSSF lepton pairs. These categories are further divided, depending on whether or not a lepton pair has $M(\ell\ell)$ in the range 75--105\GeV, consistent with a $\Z$ boson decay. Same-flavor dilepton pairs consistent with low-mass resonances are excluded from the search region with a requirement of $M(\ell\ell) > 12$\GeV.

\subsection{All-hadronic channel}\label{sec:evtreco_bHbH}

The final channel contributing to the search for \bprime{} quarks includes events reconstructed in an all-hadronic topology, to increase sensitivity to the $\PQb\PH$ decay mode of the \bprime{} vectorlike quark.  The search in this channel is designed for Higgs bosons decaying to a pair of \PQb quarks.  Because of the high mass of the \bprime{} quark, the Higgs boson is expected to be highly Lorentz boosted; consequently the \PQb quarks from the Higgs boson decay have a small angular separation. To reconstruct this signature, jet substructure algorithms are used.  The Higgs boson is reconstructed using a single CA8 jet.  This jet is required to have $\pt > 300$\GeV. The pruned jet mass is required to be in the range $90 < M({\text{jet}}) < 140$\GeV, to be consistent with the Higgs boson mass.  The $N$-subjettiness observables $\tau_2$ and $\tau_1$ \cite{Thaler:2010tr} are used to further increase the purity of events containing the two-prong decay of the Higgs boson in the \Hbb decay mode. We require the condition $\tau_2 / \tau_1 < 0.5$ to ensure that jets containing two distinct deposits of energy (subjets) are selected as the \Hbb candidates. Finally, the two identified subjets are required to be individually \PQb tagged using the CSV algorithm.  Jets satisfying all of these criteria are known as \PH-tagged jets. At least one reconstructed \PH-tagged jet is required for the final event selection.

Events are also required to have at least one additional \PQb-tagged jet, to reconstruct the \cPqb{} quark originating directly from the \bprime{} quark decay. Events are categorized according to the number of \PQb-tagged jets: exactly one or at least two. To further reduce background contributions to the event selection, a requirement is made on \HT, defined for this channel as the scalar \pt sum of all AK5 jets with \pt above 50\GeV. A requirement of $\HT > 950$\GeV maintains a high signal sensitivity while eliminating most of the multijet background.

\section{Estimation of backgrounds}\label{sec:bkgestimation}

In this section we describe the variety of the methods used to estimate the background contributions for each of the channels contributing to the search.  Detailed descriptions of the systematic uncertainties applied to these methods, and shown in the figures presented here, can be found in Sec.~\ref{sec:systematics}.

\subsection{Lepton+jets channel}\label{sec:bkgestimation_ljets}

The dominant background contribution to the lepton+jets analysis is SM top quark pair production, accounting for 77\% of the expected background yield. Other processes also contribute, including $\PW/\Z$+jets, single top quark, diboson, and \ttbar plus vector boson production, which together account for 17\% of the expected background yield.  The electroweak backgrounds are taken from simulation. The remaining contribution to the background estimation is due to multijet events. To model and estimate the contribution from these processes, control samples in data are used. The \ST  shape is taken from a multijet-enriched region defined by the selection of nonisolated leptons, or in the case of electrons, those failing the identification criteria. The \ST shapes from nonisolated leptons and isolated leptons were compared for several kinematic selections in both channels and were found to be consistent. The normalization is obtained by fitting the $\ETslash$ distribution in data individually in the 0, 1, and $\geq 2$ boson categories. The electroweak backgrounds are constrained to their expected cross sections and allowed to float within uncertainties, while the multijet normalization is allowed to float freely in these fits.

\begin{figure*}[!htbp]
\centering
\includegraphics[width=0.48\textwidth]{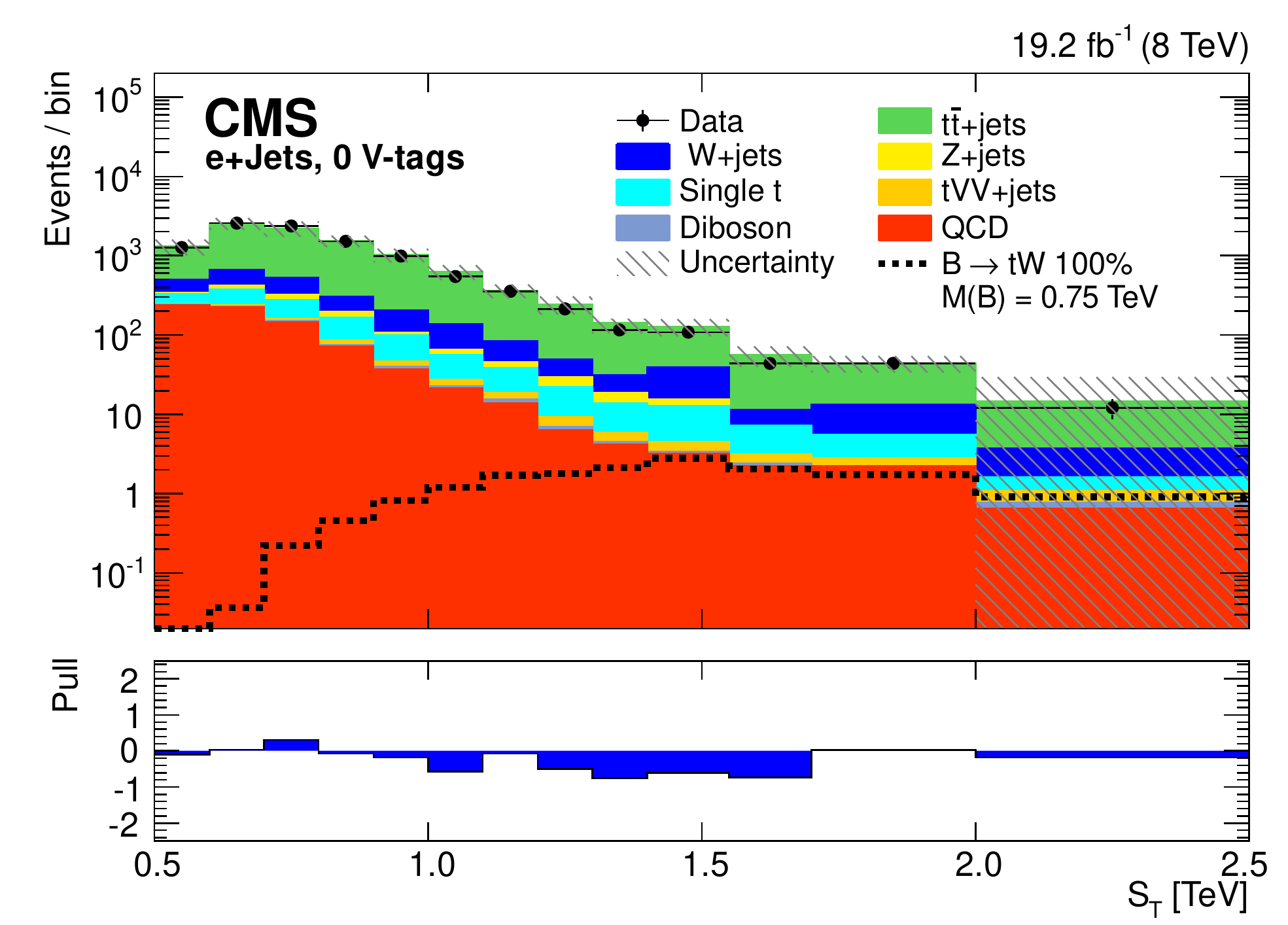}
\includegraphics[width=0.48\textwidth]{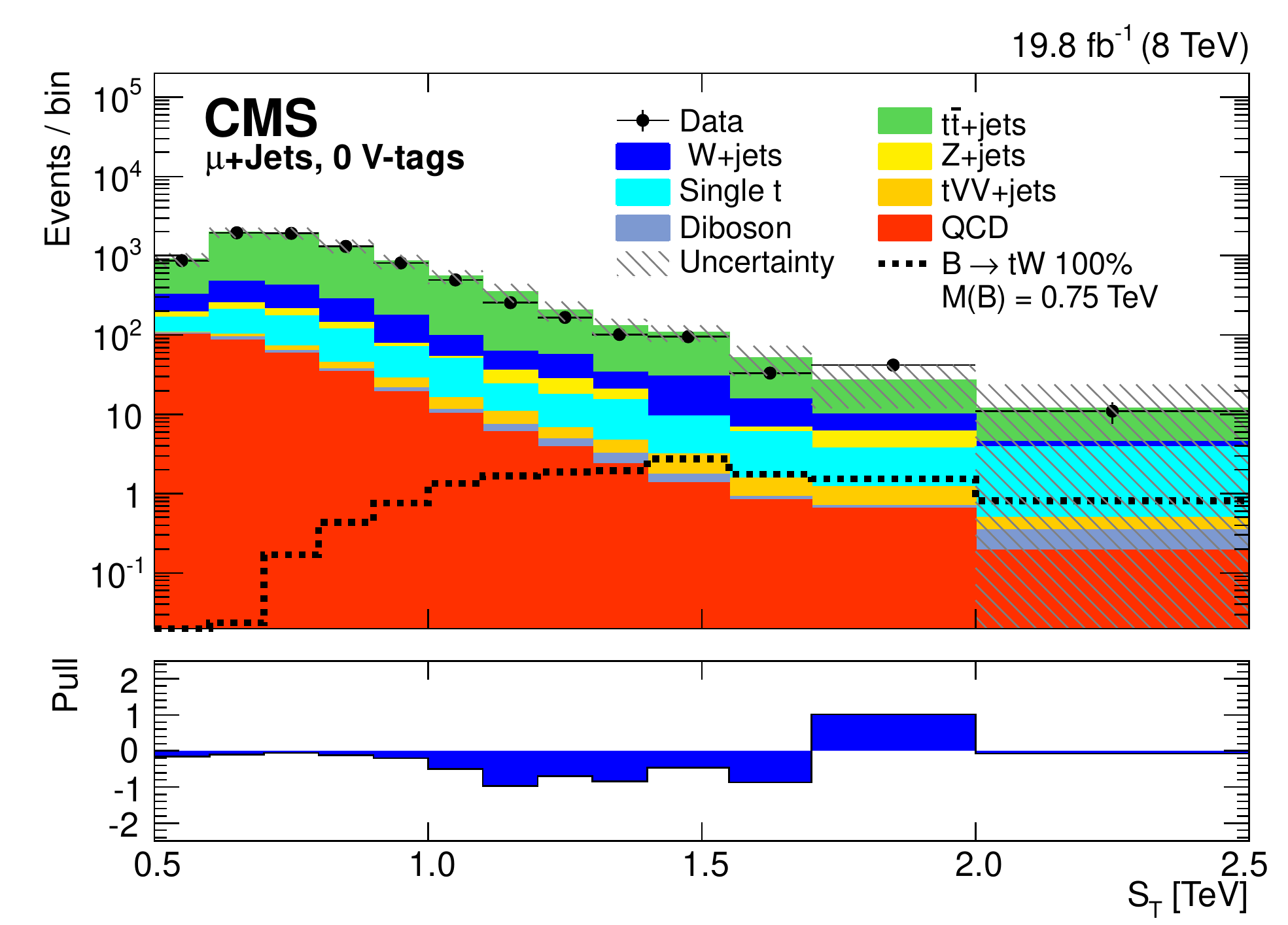}
\includegraphics[width=0.48\textwidth]{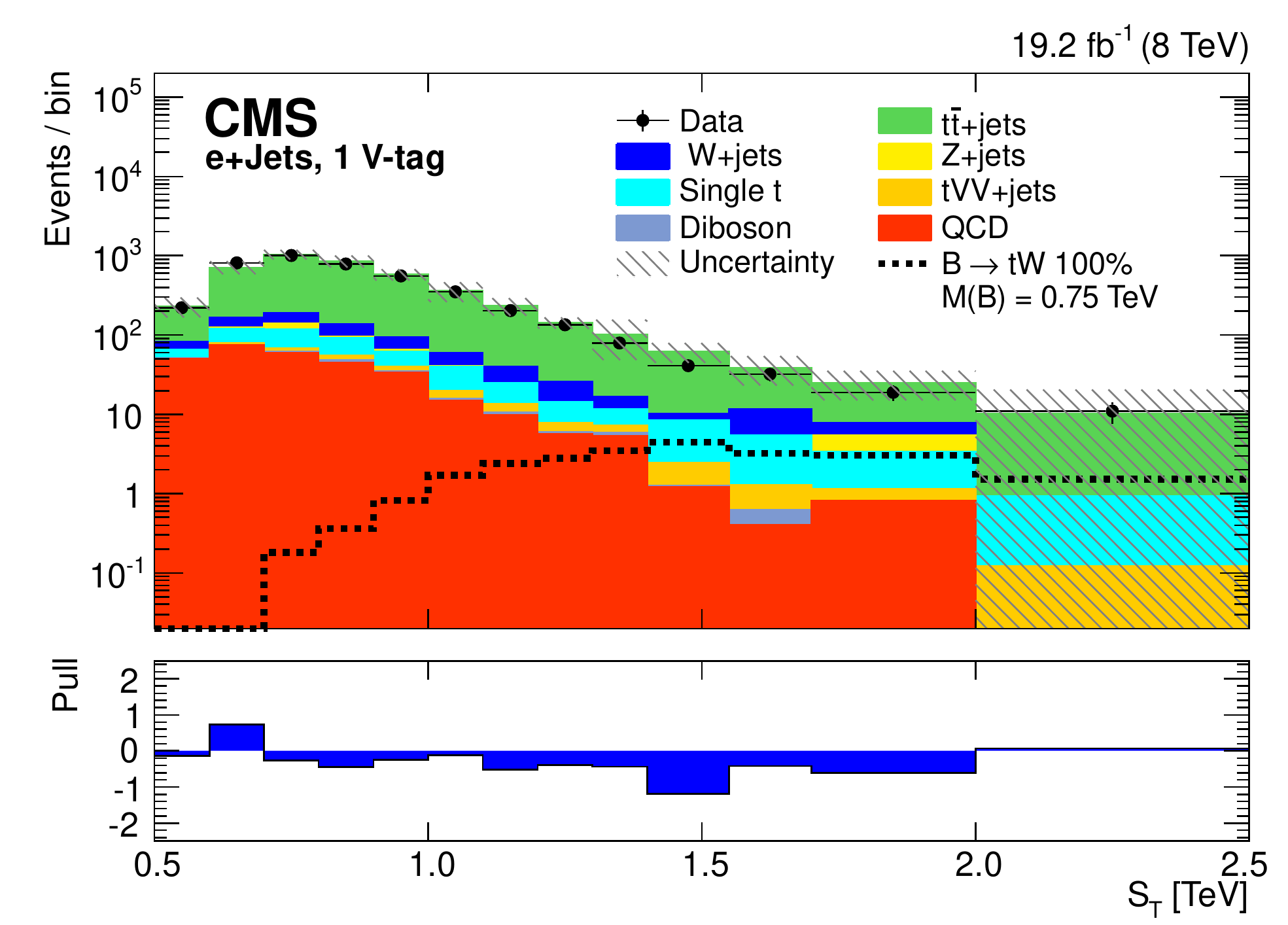}
\includegraphics[width=0.48\textwidth]{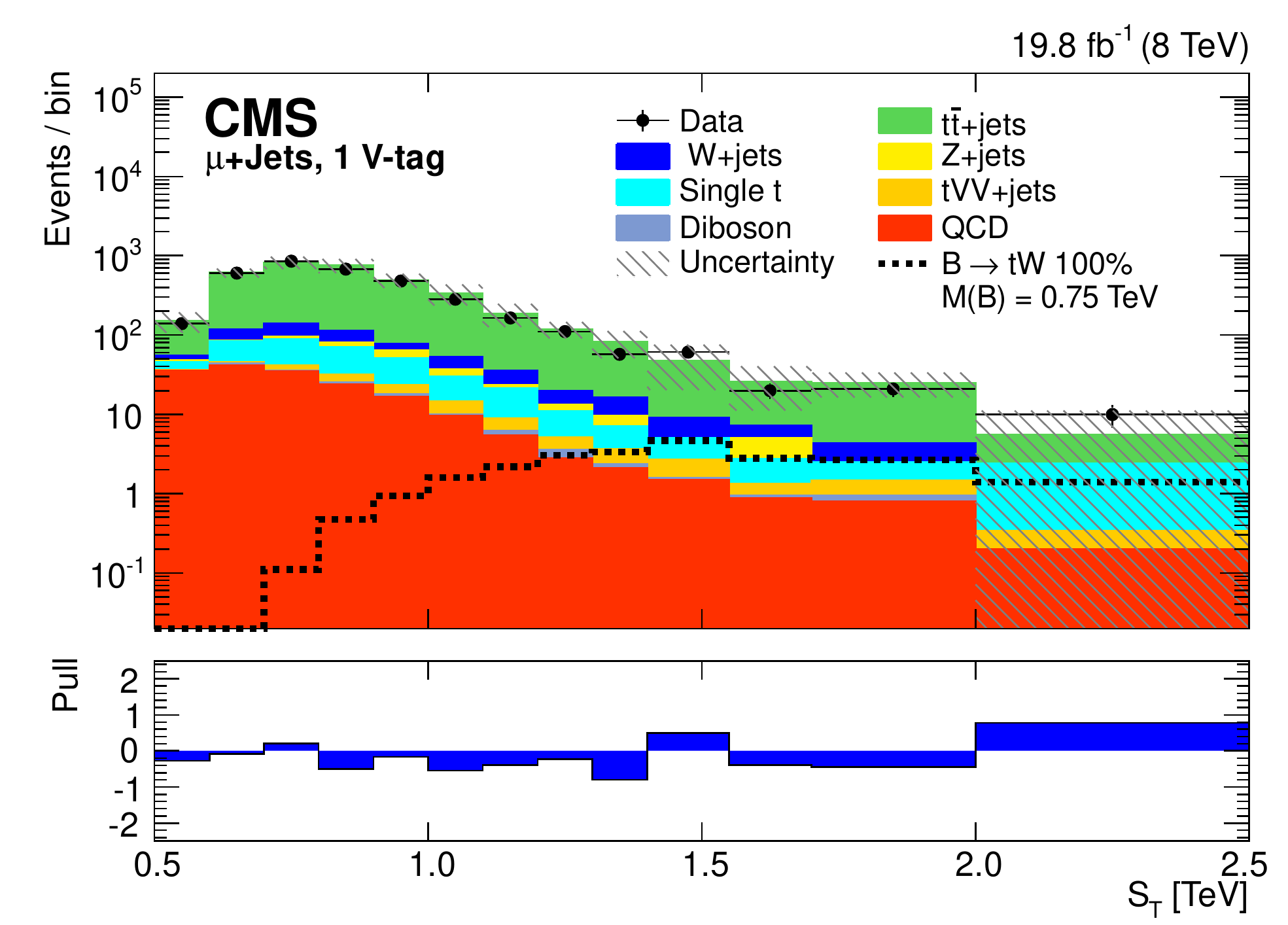}
\includegraphics[width=0.48\textwidth]{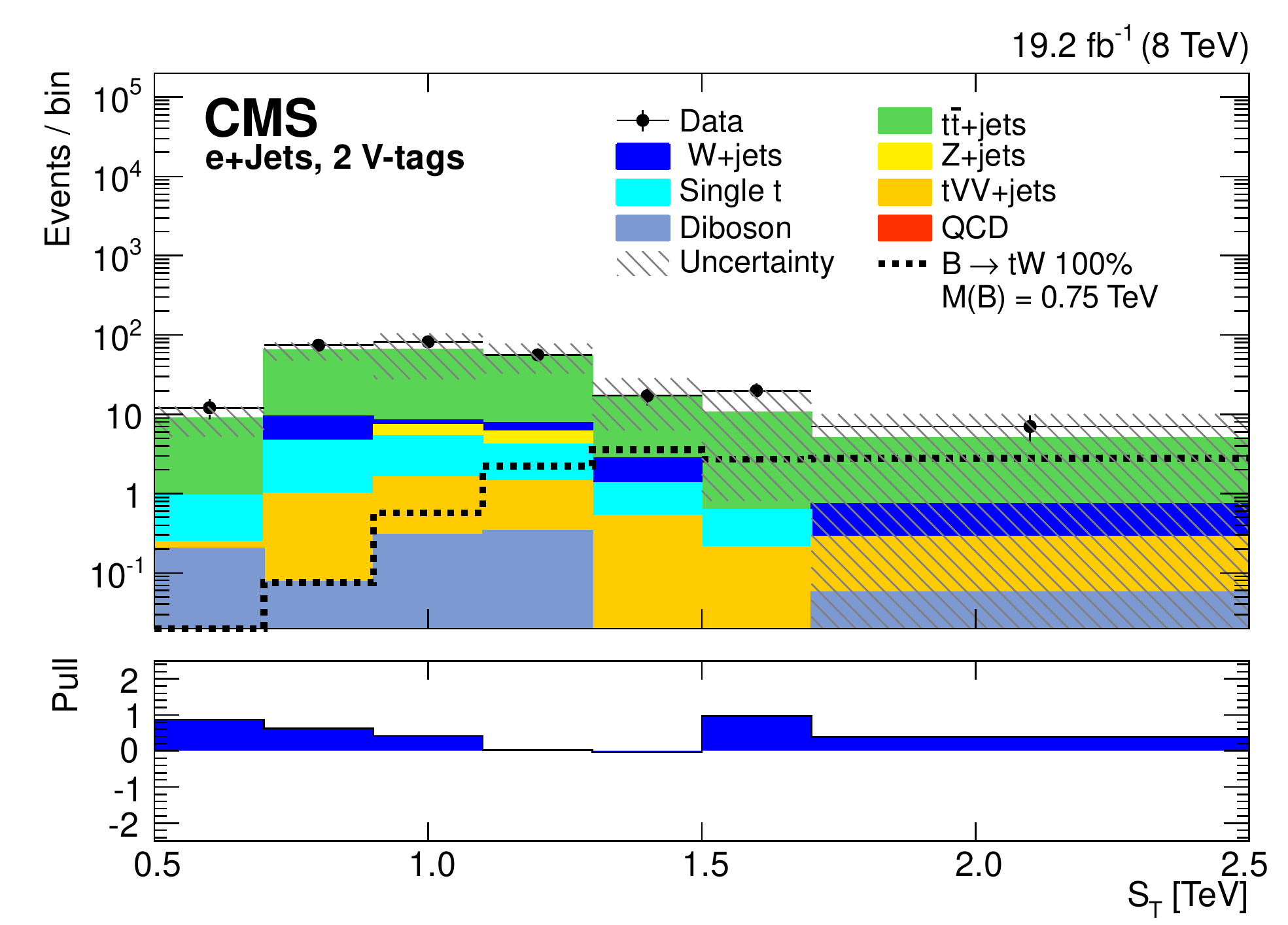}
\includegraphics[width=0.48\textwidth]{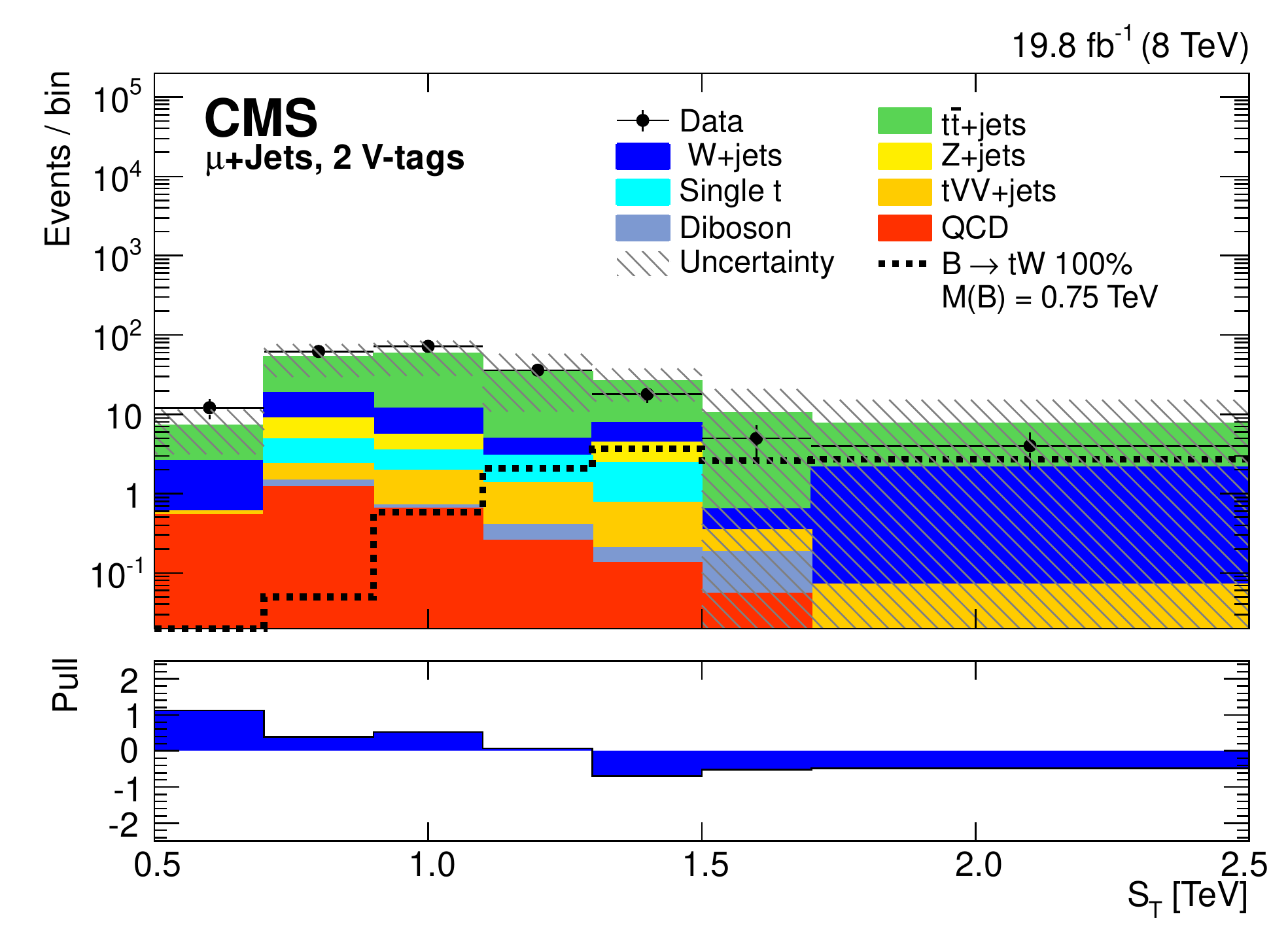}
  \caption{The $\ST$ distributions in the 0, 1, and $\geq 2$ $V$-tag categories in the electron+jets channel (left) and muon+jets channel (right). The uncertainty bands shown include statistical and all systematic uncertainties, added in quadrature for each single bin.  The horizontal bars on the data points indicate the bin width. The difference between the observed and expected events divided by the total statistical and systematic uncertainty of the background prediction (pull) is shown for each bin in the lower panels.}
  \label{fig:ljets_ST}
\end{figure*}

Events are categorized based on the flavor of the identified lepton, as well as the number of identified heavy-boson-tagged jets (V tags), including 0, 1, and  $\geq$2.  Figure~\ref{fig:ljets_ST} shows the \ST  distributions for these categories, which are used for signal discrimination in this channel.

\subsection{Same-sign lepton pair channel}\label{sec:bkgestimation_ss2l}

The background contributions for the dilepton channel with same-sign leptons are divided into distinct categories.  The first category includes events with two prompt leptons having the same charge.  This category of events represents an irreducible background composed of various SM processes, including $\ttbar\PW$, $\ttbar\Z$, diboson, and triboson production.  These background contributions are modeled using simulated events.

A second category of background events arises when the charge of one of the leptons from an oppositely charged pair is mismeasured, which happens most frequently in the same-sign $\Pe\Pe$ channel.  To model this contribution, the charge misidentification rate is measured using a control sample enhanced in \Z{}+jets and \ttbar events having two leptons in the final state.  The charge misidentification rate is extracted from the ratio of the number of events in this selection having same-sign lepton pairs to the number of events having those of opposite sign.  This contribution to the background model is then normalized by selecting opposite-sign lepton pairs in the signal region and multiplying by this charge misidentification rate.

Finally, there can be events passing the selection containing either one or two nonprompt leptons that pass the analysis lepton criteria.  To estimate this background contribution a looser lepton selection is applied, where the isolation requirement is relaxed for electrons; the isolation, impact parameters, and track quality requirements are relaxed for muons.  Leptons passing these relaxed criteria are known as ``loose'' leptons, while those passing the signal region selection are known as ``tight'' leptons.  Using data events, misidentification rates for nonprompt leptons to be reconstructed as tight leptons are measured ($f_i$, where $i$ is 1 for the leading and 2 for the subleading nonprompt lepton), along with the rates for prompt leptons to be reconstructed as tight leptons ($p_i$, where $i$ is the index of the prompt lepton in this case).  Using these loosened selection criteria, the expected yields for events containing 0, 1, or 2 nonprompt leptons ($N_{ff}$, $N_{pf/fp}$, $N_{pp}$, respectively, where the subscript $f$ refers to misidentified leptons) can be computed by using the observed numbers of events containing 0, 1, or 2 loose leptons ($N_{TT}$, $N_{TL/LT}$, $N_{LL}$, respectively), according to the relation shown in Eq.~(\ref{eqn:ssdl_matrix}):
\begin{multline}
\begin{pmatrix}  N_{pp}& N_{pf}&  N_{fp}& N_{ff} \end{pmatrix}
= \begin{pmatrix}  N_{LL}&                N_{TL}&                N_{LT}&                N_{TT}
\end{pmatrix}\\
\times\begin{pmatrix}
(1-p_{1})(1-p_{2})&p_{1}(1-p_{2})&(1-p_{1})p_{2}&p_{1}p_{2}\\
(1-p_{1})(1-f_{2})&p_{1}(1-f_{2})&(1-p_{1})f_{2}&p_{1}f_{2}\\
(1-f_{1})(1-p_{2})&f_{1}(1-p_{2})&(1-f_{1})p_{2}&f_{1}p_{2}\\
(1-f_{1})(1-f_{2})&f_{1}(1-f_{2})&(1-f_{1})f_{2}&f_{1}f_{2}
\end{pmatrix} ^{-1}.
\label{eqn:ssdl_matrix}
\end{multline}

\begin{figure*}
\centering
\includegraphics[width=0.48\textwidth]{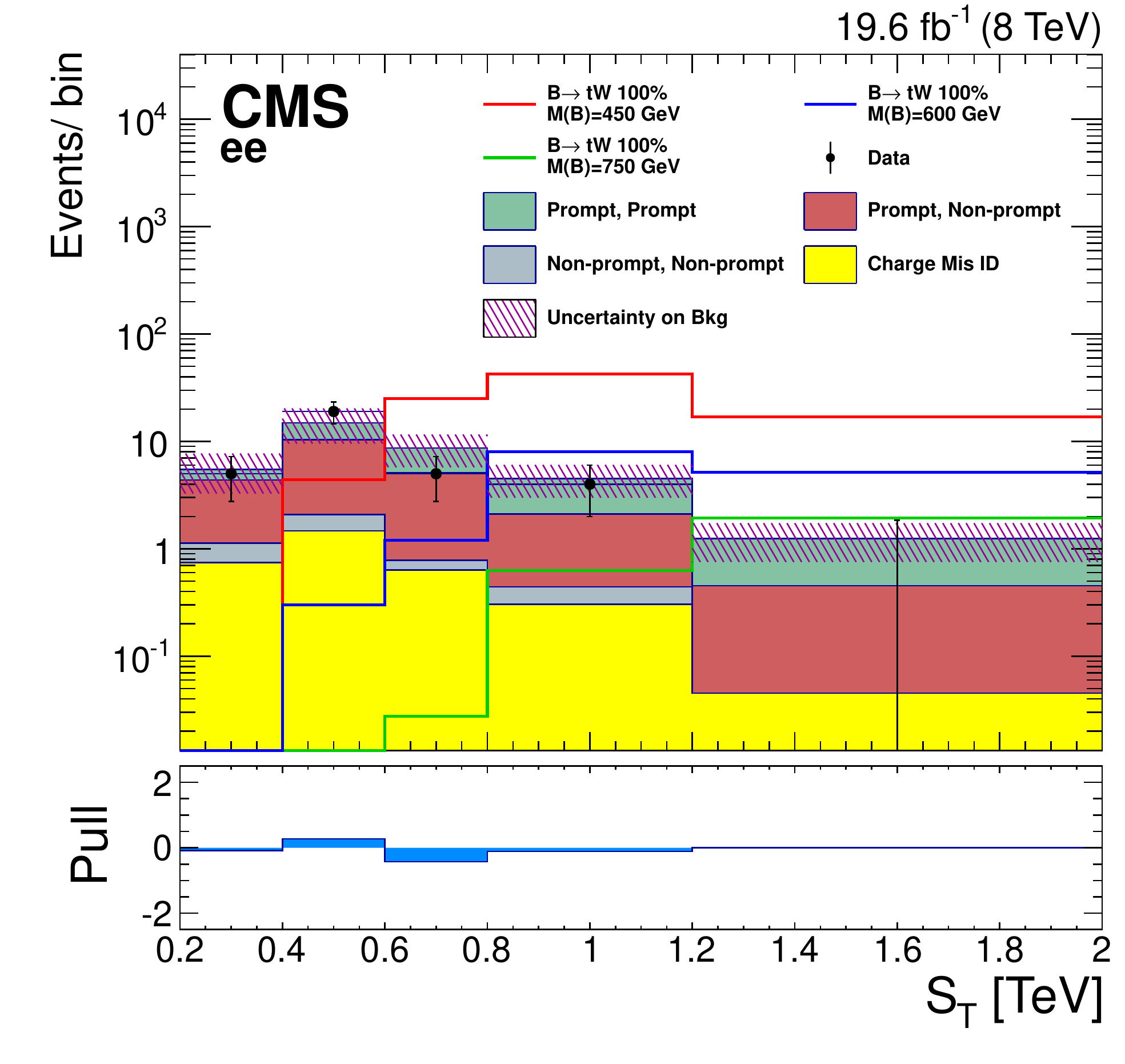}
\includegraphics[width=0.48\textwidth]{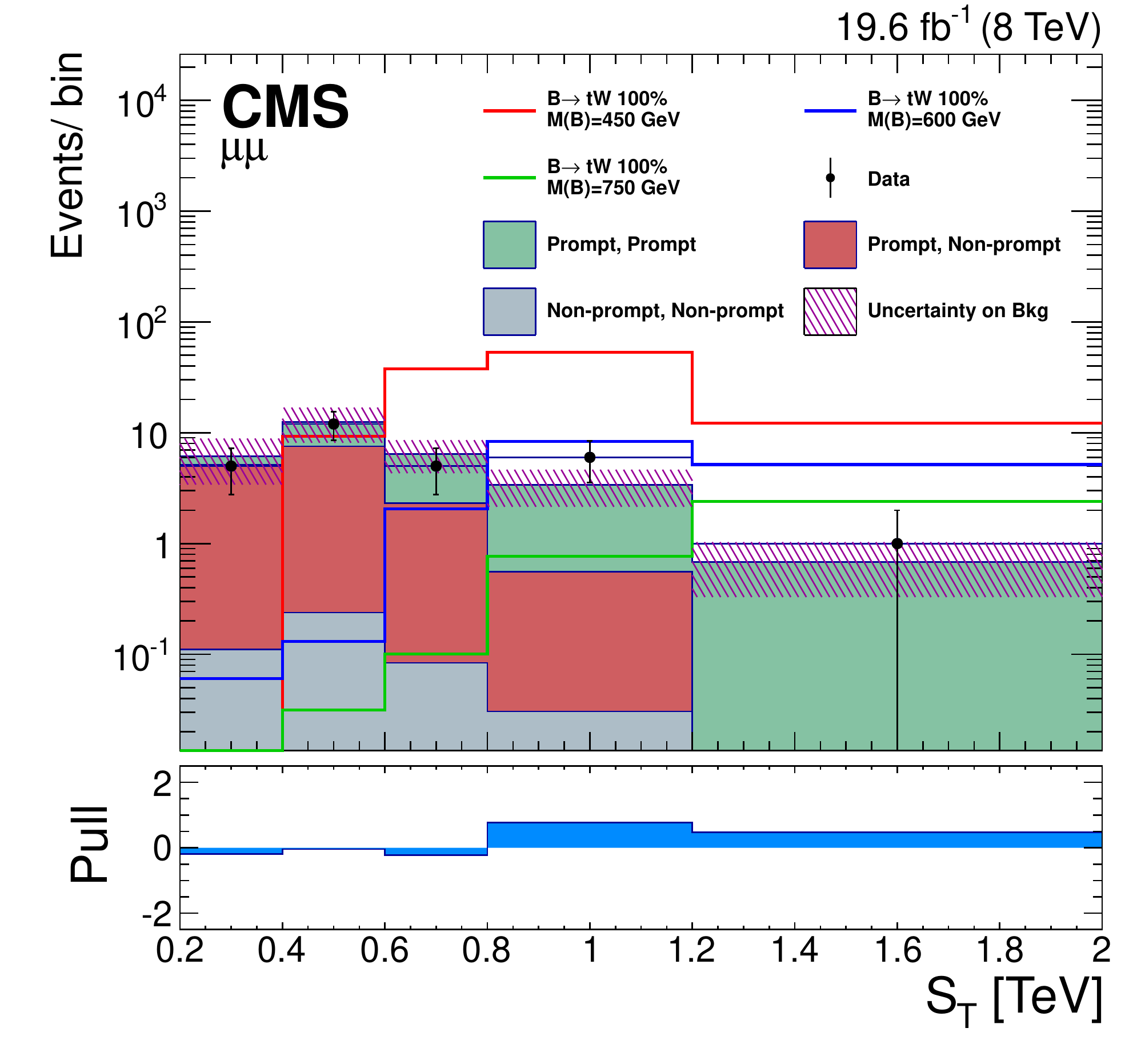}
\includegraphics[width=0.48\textwidth]{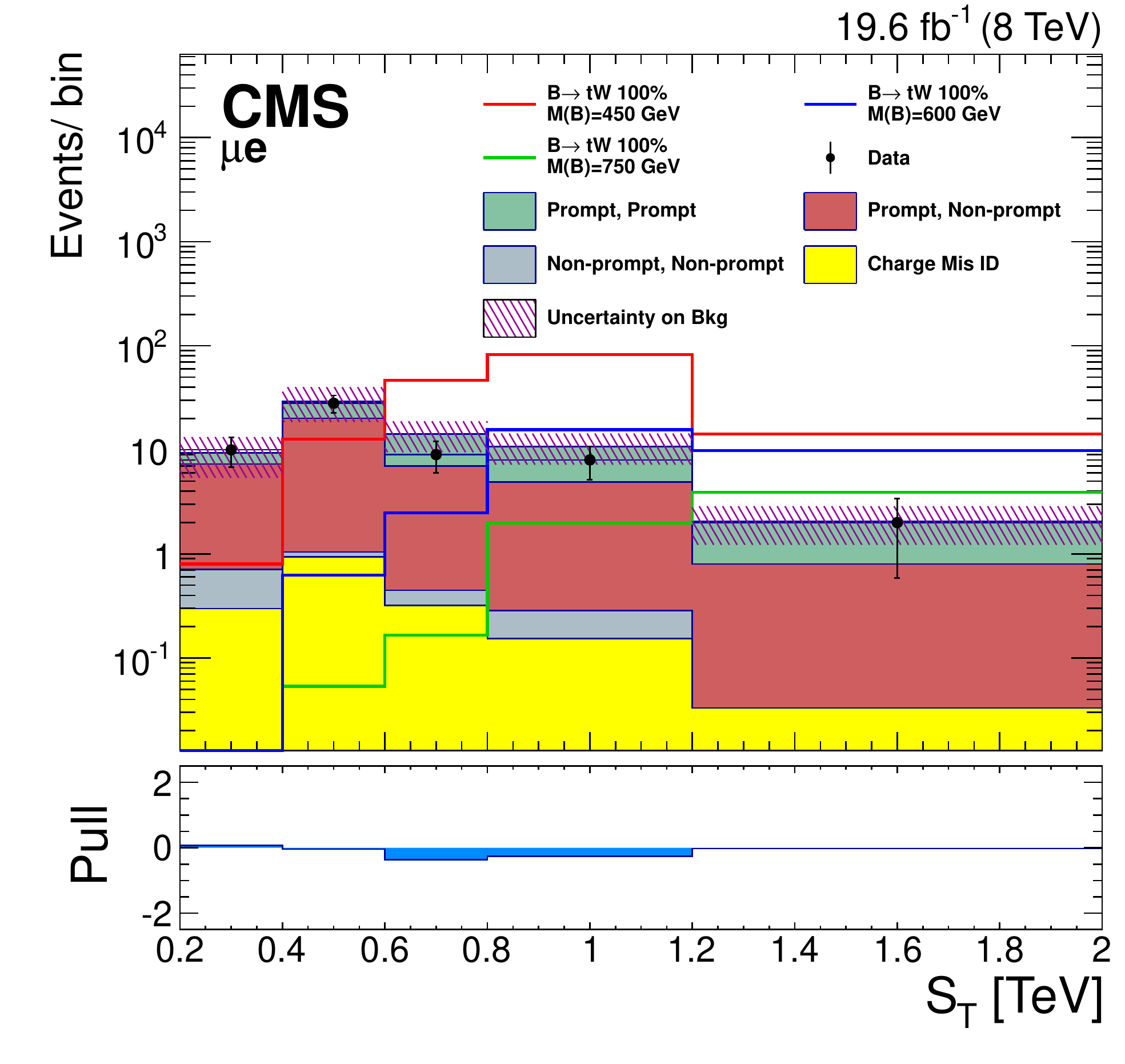}
\caption{The \ST  distributions used for signal discrimination in the same-sign dilepton channel.  The distribution is shown for the three dilepton categories used: dielectron (top left), dimuon (top right), and electron-muon  (bottom). The horizontal bars on the data points indicate the bin width. The difference between the observed and expected events divided by the total statistical and systematic uncertainty of the background prediction (pull) is shown for each bin in the lower panels.  }
\label{fig:ssdil_ST}
\end{figure*}

After using this method to estimate the background from the contributions containing nonprompt leptons, the \ST  distribution is used to discriminate signal events from background.  The \ST  distributions for the dielectron, dimuon, and electron-muon channels are shown in Fig.\ \ref{fig:ssdil_ST}.

\subsection{Opposite-sign lepton pair channel}\label{sec:bkgestimation_os2l}

The main background in the opposite-sign dilepton channel is from the inclusive \Z{}+jets process (93\%), with the remaining fraction due to \ttjets and diboson processes. Instead of using simulated events, control samples in data are used to predict the normalization and shape of the \mellellb spectrum of the background. The background is estimated from data using an $ABCD$ method to predict the $\cPqb\Z$ invariant mass distribution \mellellb in the signal region, labeled $B$, using control regions $A$, $C$, and $D$. The classification of the events into region $A$, $B$, $C$, or $D$ is made using event selection variables that are largely uncorrelated for the background samples. The two variables chosen are the number of jets, $N_\text{jets}$, and the \PQb-tagging discriminator of the highest \pt jet in the event.  With an identified $\Z$ boson decaying leptonically, there will be at least two jets expected in signal events, providing discrimination power against SM background processes.

\begin{figure}[!Hhtb]
 \centering
  \includegraphics[width=0.45\textwidth]{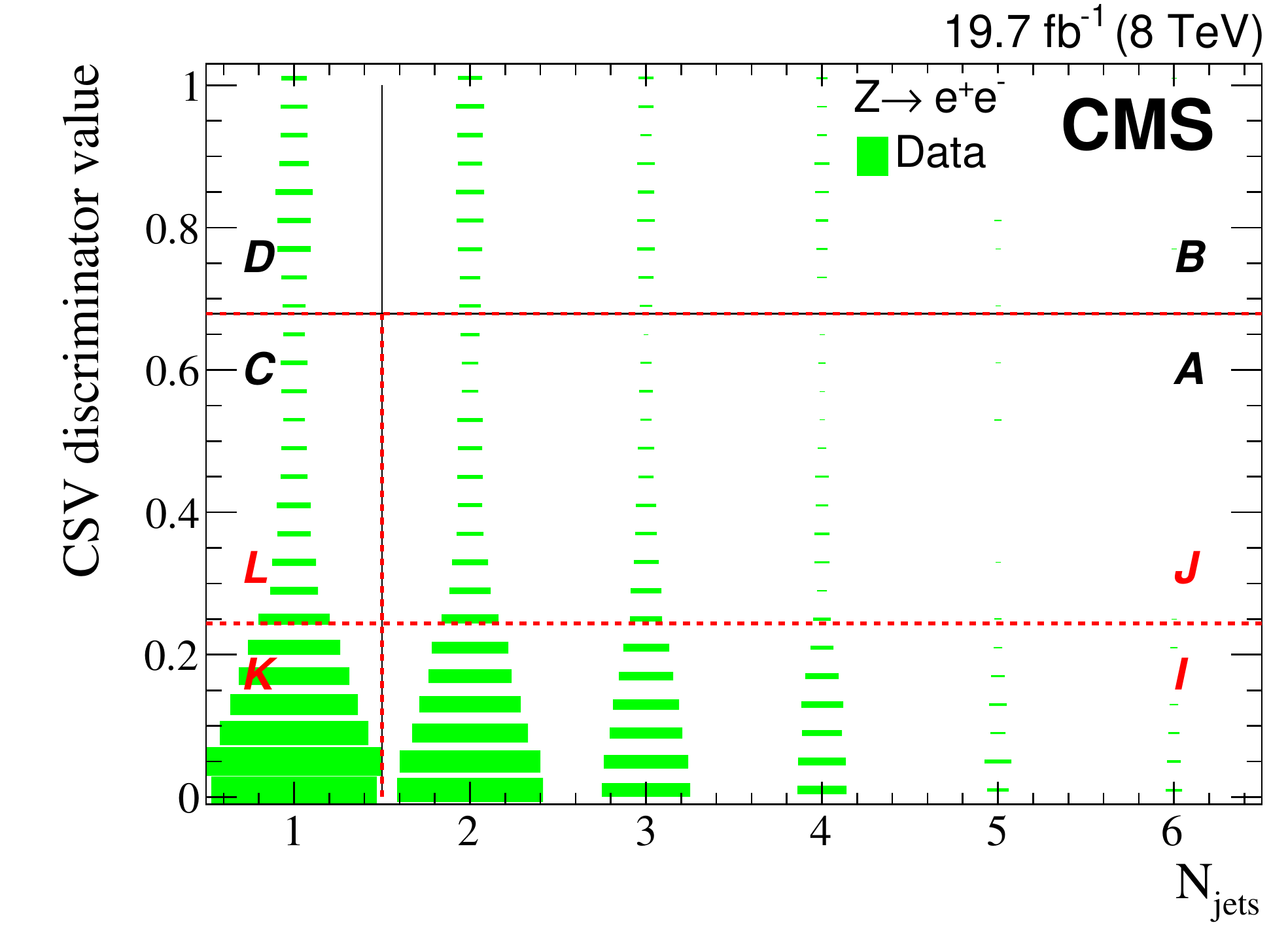}
  \includegraphics[width=0.45\textwidth]{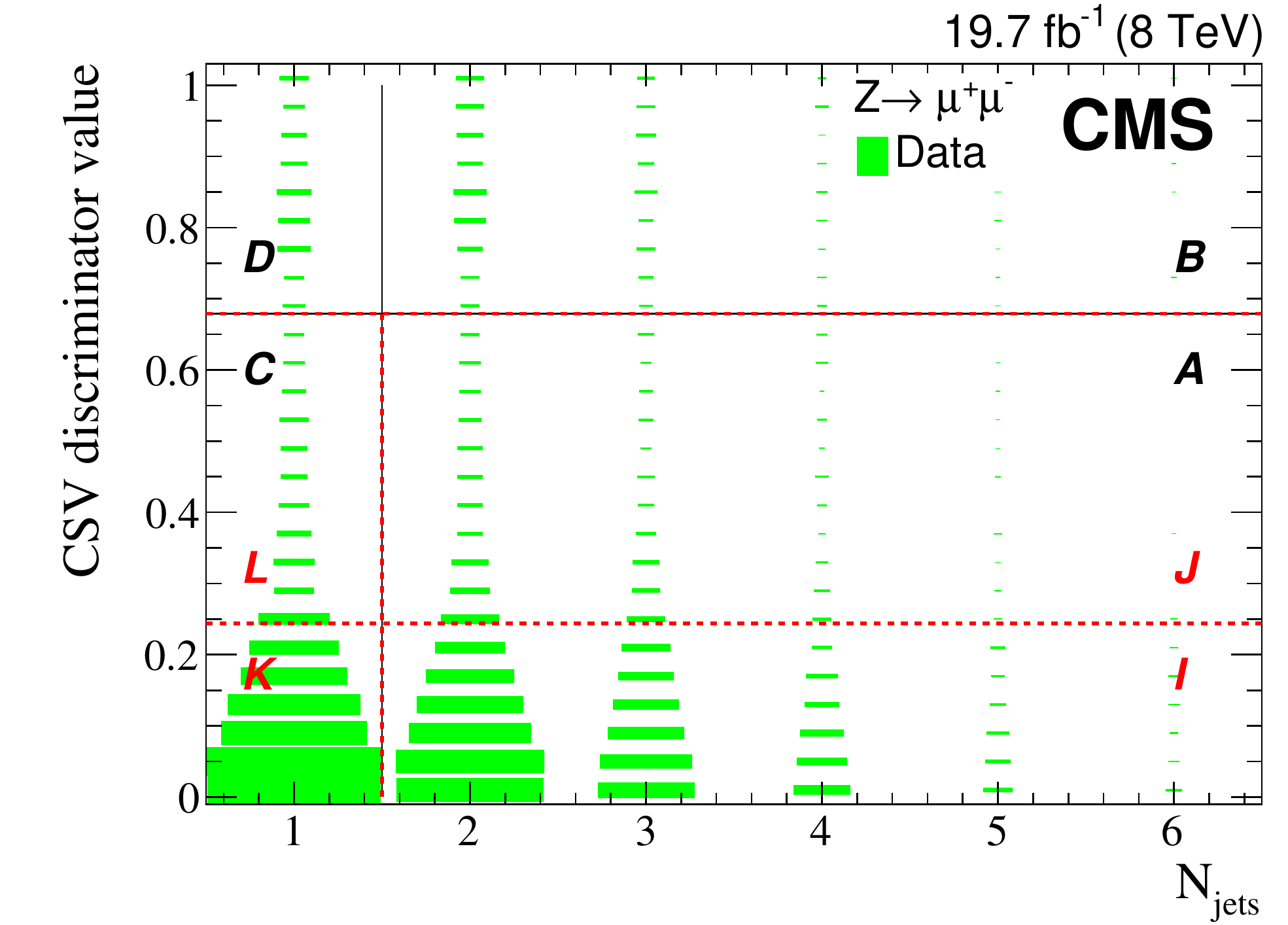}
  \caption{The event distribution in the plane of $N_\text{jets}$ vs the \PQb-tagging discriminator value, used to define the regions $A$, $B$, $C$, and $D$ for the opposite-sign dilepton \Zelel (\cmsLeft) and \Zmumu (\cmsRight) channels.    The region $B$ is the signal region while the others constitute the control regions.  The regions $I$, $J$, $K$, and $L$ are used for estimation of systematic uncertainties.  All other selection criteria used to select the \bprime{} quark candidates have been applied. The area of each bar is proportional to the number of events in a given bin of the distribution of $N_\text{jets}$ vs \PQb-tagging discriminator.}
  \label{fig:abcd}

\end{figure}

The selections used are ($i$) either $N_\text{jets} = 1$ or $N_\text{jets} > 1$ and ($ii$) events with the leading jet either passing or failing the \PQb-tagging discriminator threshold ($>$0.679). These selections divide the $N_\text{jets}$ vs. \PQb-tagging discriminator plane into the four regions shown in Fig.~\ref{fig:abcd}. The signal contribution outside the signal region $B$ was found to be negligible using simulated event samples. Under the hypothesis of complete noncorrelation between $N_\text{jets}$ and the \PQb-tagging discriminator, the number of background events in the signal region would be given by
\begin{math}
N_{B} = N_{A} \times N_{D}/N_{C},
\end{math}
where $N_{X}$ is the number of events in the corresponding region. However, residual correlation between the two variables is present and must be taken into account in the background estimation procedure. The correlation is measured from data using an alternative set of control regions defined using the following criteria:
($i$) $N_\text{jets} = 1$ or $N_\text{jets} > 1$ and
($ii$) $0.244 < \PQb\text{-tagging discriminator} < 0.679$ or $\PQb\text{-tagging discriminator} < 0.244$ for the leading jet.
This classification divides the $N_\text{jets}$ vs the \PQb-tagging discriminator plane into four regions, labeled $I$, $J$, $K$, and $L$, as shown in Fig.~\ref{fig:abcd}. These four regions are completely contained within the previously defined regions $A$ and $C$.
The ratio
\begin{math}
{\cal C} = N_{J} N_{K}/N_{I} N_{L}
\end{math}
is equal to 1 if $N_\text{jets}$ and the \PQb-tagging discriminator variables are perfectly uncorrelated, and is used to quantify the degree of correlation between the two. The number of background events, taking into account the correlations, is given by $N_{B}\,\mathcal{C}$. The values of $\mathcal{C}$ were measured to be $1.29\pm0.08$ for \Zelel and $1.38\pm0.07$ for the \Zmumu channels, where the uncertainties are statistical and related to the sample sizes in the regions $I$, $J$, $K$, and $L$. These factors are significantly different from unity, implying some degree of correlation between $N_\text{jets}$ and the \PQb-tagging discriminator. Closure tests were performed with simulated samples, as well as with data control samples with selections orthogonal to those for the regions described above. The values of the correlation factors obtained were consistent within uncertainties and stable with respect to the variation of the $\PQb\text{-tagging discriminator}$ values within $\pm 10\%$.

While the above procedure is used to predict the total number of background events, the shape of the \mellellb background distribution is assumed to be the same in the signal region and the region $A$. This assumption is justified by examining the \mellellb distributions in the signal region and in region $A$, using simulated events.  The shapes obtained are consistent within the uncertainties in each.   The total event yields in data and the estimated background are given in Table~\ref{tab:FinalEvtYields_8TeV}. The uncertainty in this background estimation is given by a combination of the statistical and systematic sources described in Sec.~\ref{sec:systematics}.

\begin{table}[!Hhtb]
\topcaption{
Expected background yields and observed number of events in data in the opposite-sign dilepton channel. The background is obtained from data.  The background uncertainties include both statistical and systematic components.
}
\label{tab:FinalEvtYields_8TeV}
\centering
\begin{scotch}{lcc}
& \Zelel & \Zmumu \\
\hline
Expected background & $379 \pm 70$  & $534 \pm 79$ \\

Observed events      & $334$          & $542$         \\
\end{scotch}

\end{table}

\begin{table*}[!Hhtb]
\topcaption{Expected signal event yields in the opposite-sign dilepton channel, shown for \bprime{} quark masses $M(\bprime)$ from 450 to 800\GeV and for two values of the branching fraction.}
\label{tab:FinalEvtYields325_Signal_BR100_BR50}
\centering
\begin{scotch}{cxxxx}
 & \multicolumn{2}{c}{$\mathcal{B}(\bprimetobZ) = 100\%$} & \multicolumn{2}{c}{$\mathcal{B}(\bprimetobZ) = 50\%$} \\ \cline{1-5}
\multicolumn{1}{c}{$M(\bprime)$ [\GeVns{}]}          & \multicolumn{1}{c}{\Zelel} & \multicolumn{1}{c}{\Zmumu} & \multicolumn{1}{c}{\Zelel} & \multicolumn{1}{c}{\Zmumu} \\[\cmsTabSkip]
450  & 214,13  & 336,16  & 102,4     & 162,5    \\
500  & 122,7   & 209,9   & 56,2     & 94,3    \\
550  & 76,4   & 114,5   & 33,1     & 54,2    \\
600  & 36,2   & 66,3   & 17.6,0.7   & 30.8,0.9  \\
650  & 23,1   & 41,2   & 11.0,0.4   & 19.5,0.6  \\
700  & 14.1,0.7 & 25.9,1.0 & 6.5,0.2   & 12.0,0.3  \\
750  & 7.6,0.4 & 15.5,0.6 & 3.6,0.1   & 7.4,0.2  \\
800  & 4.8,0.3 & 9.9,0.4 & 2.2,0.1  & 4.6,0.1  \\
\end{scotch}
\end{table*}

\begin{figure}[!htb]
 \centering
  \includegraphics[width=0.45\textwidth]{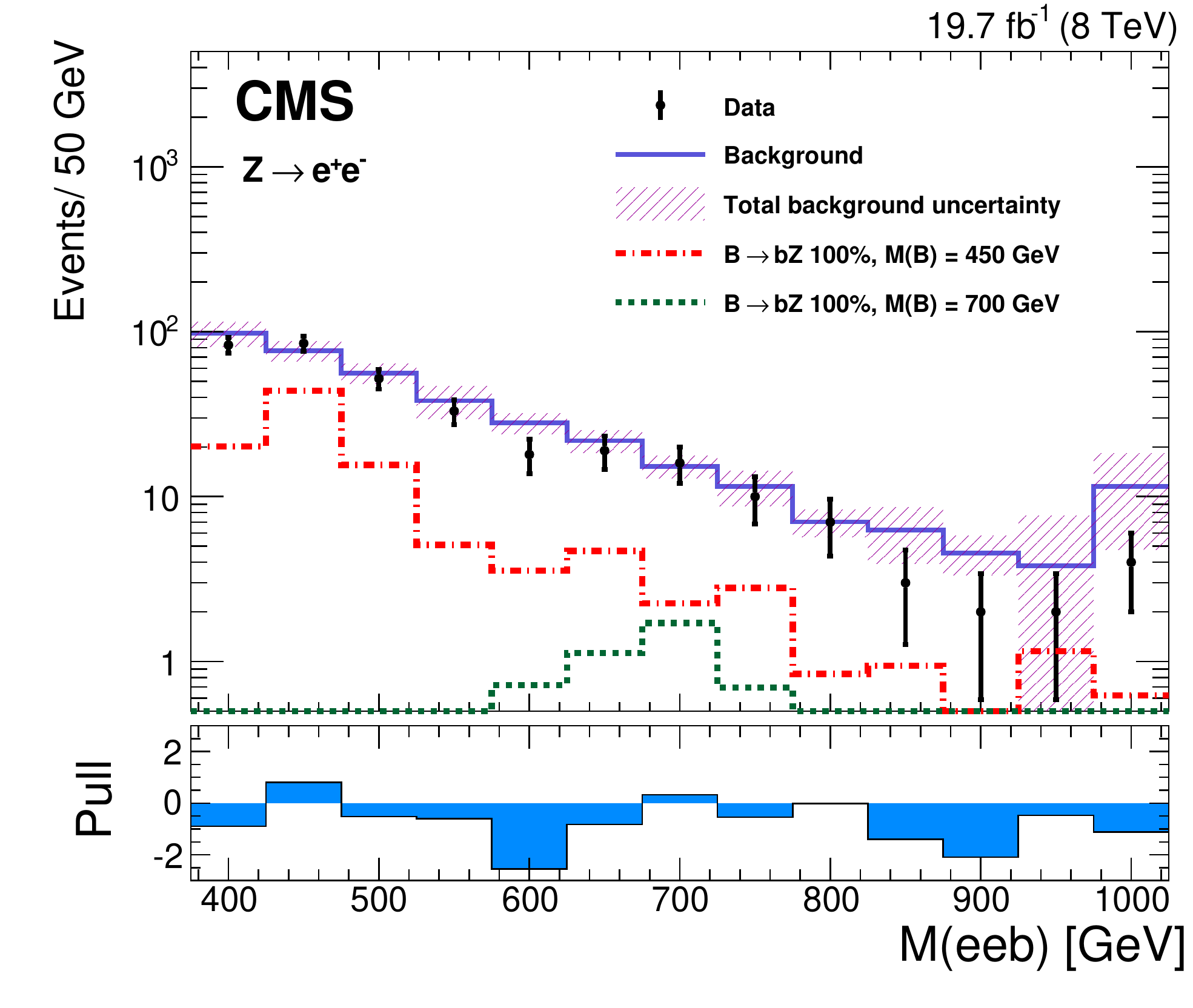}
  \includegraphics[width=0.45\textwidth]{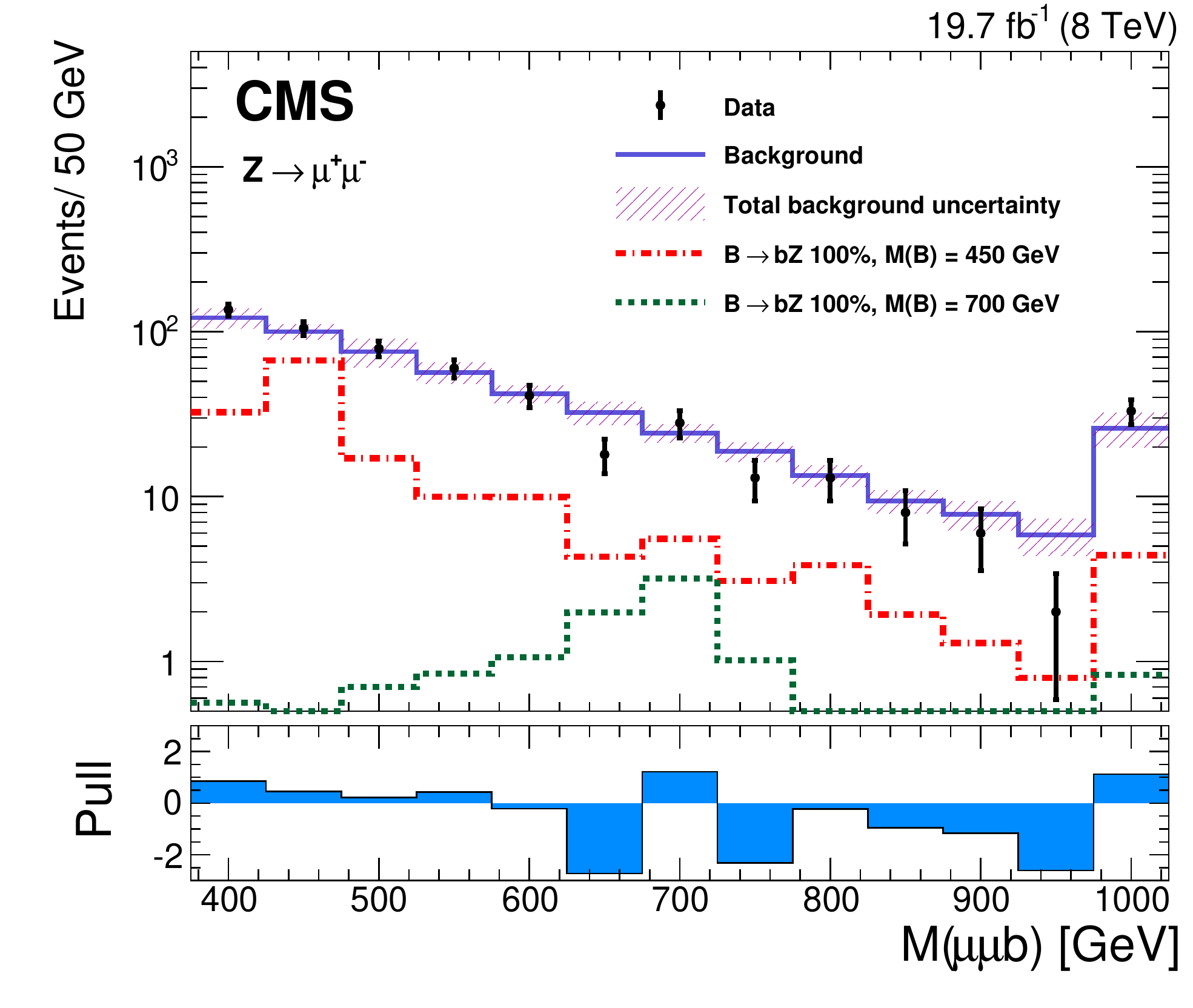}
  \caption{The invariant mass of reconstructed \bprime{} quark candidates in the opposite-sign dilepton \Zelel (\cmsLeft) and \Zmumu (\cmsRight) channels. The estimated background is shown by the solid line, along with the total uncertainty (hatched area). The last bin of the histograms contains all events with $\mellellb > 1000$\GeV. The signal contribution is shown for two \bprime{} quark masses. The difference between the observed and expected events divided by the total statistical and systematic uncertainty of the background prediction (pull) is shown for each bin in the lower panels.}
  \label{fig:bkgDataDriven_8TeV}
\end{figure}

The expected yields for the signal with different \bprime{} quark masses and two different values of the branching fraction, 100\% and 50\%, for \bprimetobZ are given in Table~\ref{tab:FinalEvtYields325_Signal_BR100_BR50}.
Since we require exactly one opposite-charge lepton pair, the probability of identifying a \bprime{} quark does not depend on the decay of the other \bprime{} quark.
Figure~\ref{fig:bkgDataDriven_8TeV} shows the mass spectra of the reconstructed \bprime{} quark candidates, and the estimated background. The expected \bprime{} quark signals, for $M(\bprime) = 450$ and 700\GeV, are also shown. The error bars on the expected background are due to the statistical uncertainties as well as the uncertainty from the background estimation method. The overall normalization of the background agrees with the observed number of events. The \mellellb distributions in both the \Zelel\,and the \Zmumu\,channels show some discrepancies between data and expectation in a few bins, caused by a flavor dependence in the reconstructed $M(\ell\ell j)$ distribution observed in \Z{}+jets events. The control region $A$ is more enriched in light quark flavors than the signal region $B$, leading to the observed discrepancy. A systematic uncertainty is applied to cover this effect, based on the shape differences observed between the control regions $I$ and $J$ in data. In simulation, these regions were found to have relative flavor content similar to the regions $A$ and $B$.

\subsection{Multilepton channel}\label{sec:bkgestimation_mulitlepton}

In multilepton channels, the level of the SM background contribution varies considerably across event categories. The categories with hadronic tau decays or OSSF lepton pairs suffer from larger background contributions than the others. Therefore, we improve sensitivity to new physics by separating categories with low and high background contributions. We categorize events with three leptons separately from those with four or more leptons. Events with identified \PQb jets, having higher background from \ttbar events, are classified separately.

We consider backgrounds from rare processes such as $\ttbar\PW$, $\ttbar\Z$, $\cPqt\cPaqb\Z$, where simulated events are used.  The main SM background sources in multilepton+jet events include dilepton processes such as \Z{}+jets, $VV$+jets, and \ttbar{}+jets production with a misidentified lepton that passes selection criteria, and processes containing two leptons and an additional off-shell photon that undergoes a conversion, giving another reconstructed lepton. The above backgrounds are estimated using simulated events, except for the \Z{}+jets and $\PWp\PWm$+jets backgrounds, which are estimated from data, as described below.

\begin{figure}[!htbp]
\centering
\includegraphics[width=0.48\textwidth]{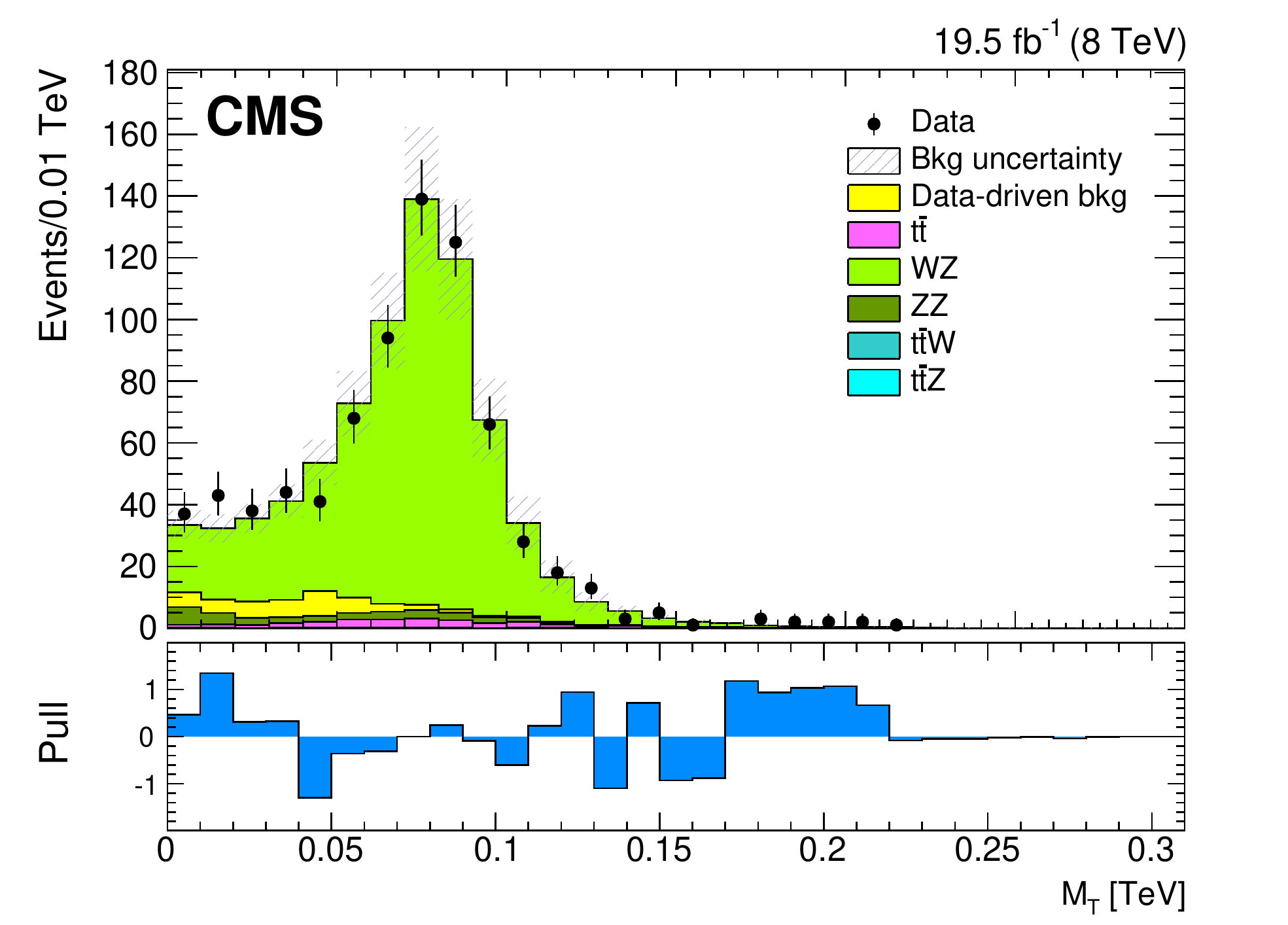}
\includegraphics[width=0.48\textwidth]{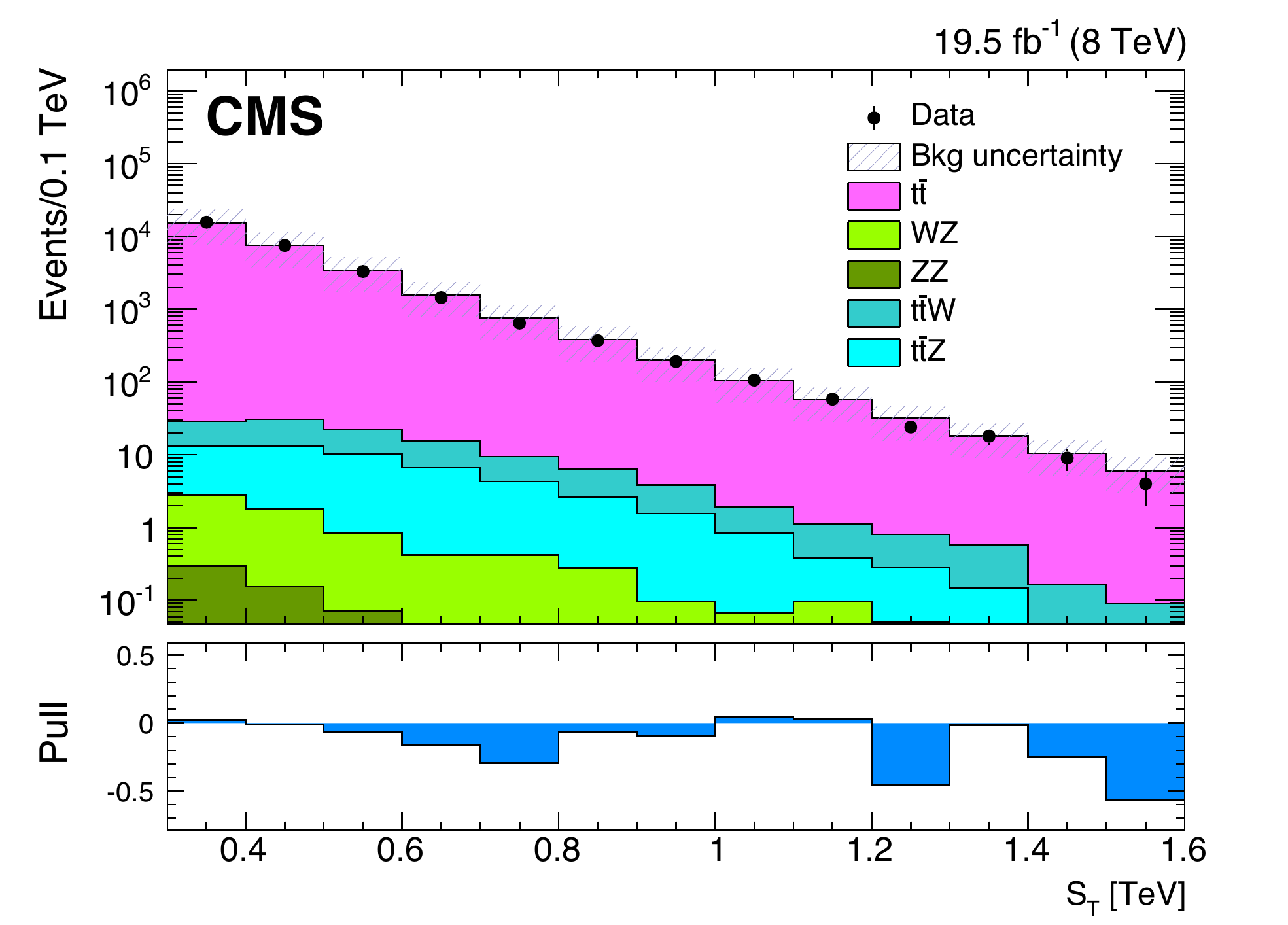}
\caption{The transverse mass $M_T$ distribution of events in a control sample of the multilepton analysis, enriched in $\PW\Z$ by requiring an OSSF pair with $M(\ell\ell)$ in the $\Z$ boson mass window and $50 <  \ETslash < 100$\GeV (\cmsLeft).  The \ST  distribution for events containing an opposite-sign $\Pe\Pgm$ pair in the \ttbar control region of the multilepton analysis (\cmsRight). Uncertainties include both statistical and systematic contributions. The difference between the observed and expected events divided by the total statistical and systematic uncertainty of the background prediction (pull) is shown for each bin in the lower panels.}
\label{fig:WZMT}
\end{figure}

Backgrounds from \ttbar enriched processes are estimated from simulation, after validation in single-lepton and dilepton control regions. In the single-lepton control region, exactly one isolated muon with  $\pt \geq 30$\GeV, at least three jets (one of which is \PQb tagged), and $\ST \geq 300$\GeV are required. The dilepton control region requires an $\Pe\Pgm$ combination and is used to compare kinematic variables such as \ST  (see Fig.~\ref{fig:WZMT}), $\HT$, and \ETslash between data and simulation.   In this channel, $\HT$ is defined as the scalar sum of selected jet \pt values.  Additionally, the distribution of the number of jets is reweighted to match data for both the single-lepton and dilepton control regions.

Standard model $\PW\Z$+jets and $\Z\Z$+jets production where both bosons decay leptonically can produce three prompt and isolated leptons with large $\HT$ and \ETslash. This class of background is irreducible since its final state cannot be distinguished from the signal scenario.  Simulated events are used to model this background contribution.
We scale the simulation to match the measured lepton efficiencies and \ETslash resolution.  We verify the simulation by comparing to a data sample enriched in $\PW\Z$ production, the dominant contribution to trilepton signatures from $VV$+jets.  The $\PW\Z$ events are selected by requiring three leptons, $50 < \ETslash <  100$\GeV, a $\Z$ boson candidate with $M(\ell^+\ell^-)$ in the range 75--105\GeV, and $\HT < 200$\GeV. We apply a constant scale factor of 1.14 to the $\PW\Z$ simulation, chosen to normalize the simulation to data in the region $50 < \ETslash < 100$\GeV for the observed transverse mass of the $\PW$ boson, shown in Fig.\ \ref{fig:WZMT}.

Off-shell photon conversions can produce a lepton with very low $\pt$ that will not pass the selection criteria or will not be reconstructed. Drell-Yan processes with such conversions can lead to a significant background for the three-lepton category. A measurement of the extrapolation factors for photon conversions to electrons or muons is performed using data events. We measure the extrapolation factors in a control region devoid of signal events, with low $\HT$ and $\ETslash$. The ratio of the number of events with  $\abs{M(\ell^+\ell^-{\ell^{\prime}}^{\pm}) - M(\Z)}  < 15\GeV$ or $\abs{M(\ell^+\ell^-{\ell}^{\pm}) - M(\Z)} < 15\GeV$ to the number of events with $\abs{M(\ell^+\ell^-{\gamma}) - M(\Z)} < 15\GeV$ defines the extrapolation factor, which is $2.1 \pm 0.3\%$ $(0.7\ \pm 0.1\%)$ for electrons (muons). The measured extrapolation factors are then used to estimate the background in the signal region from the observed number of $\ell^+\ell^- \gamma$ events in the search region. The lepton selections of this channel strongly reject external conversions, where an on-shell photon converts to an $\ell^+\ell^-$ pair in the material of the detector.

We use data to estimate background contributions from processes with two genuine leptons and one or more misidentified leptons such as $\Z(\ell\ell)$+jets and $\PWp\PWm(\ell\ell + \ETslash)$+jets. In order to estimate background from jets producing misidentified light-lepton candidates that appear to be prompt and isolated, we use data events containing two reconstructed leptons and an additional isolated track.  This contribution is then scaled by an extrapolation factor relating isolated tracks to lepton candidates from jets. This light-lepton extrapolation factor is measured in control samples where no signal is expected, such as in the low-$\ETslash$ or low-$\HT$ regions.  We measure the extrapolation factor between isolated tracks and electron (muon) candidates to be $0.7 \pm 0.2\%$ $(0.6 \pm 0.2\%)$, using a data sample dominated by \Z{}+jets. The contribution from backgrounds containing a misidentified third lepton is determined by multiplying the number of events containing isolated tracks in the sample with two leptons by the light-lepton extrapolation factor. Similarly, we estimate misidentified background contributions for the four-lepton selection by examining two-lepton events with two additional isolated tracks. Since the light-lepton misidentification rates vary with the \PQb quark content across the control samples, the rate is determined as a function of the impact parameter distribution of nonisolated tracks in data.

Unlike electrons and muons, hadronically decaying tau leptons cannot be easily identified without an isolation requirement. Therefore, the dominant background in tau identification is from jets reconstructed as $\tauh$ candidates. To measure this contribution, we loosen the isolation requirements on reconstructed tau leptons to get an extrapolation factor between nonisolated taus and isolated taus. We extrapolate the sideband region, $6 < I < 15$\GeV, to a signal region, $I < 2$\GeV, where $I$ is defined as the amount of energy reconstructed in a cone of $\Delta R < 0.3$ around the tau lepton candidate, excluding the tau lepton candidate itself. We measure the extrapolation factor for jets reconstructed as taus, defined as the ratio of the number of tau candidates in the signal region to the number in the sideband region, to be  $20 \pm 6\%$. The ratio is applied to a selection of events in a sideband region containing two light leptons and a tau lepton to estimate the contribution from misidentified $\tauh$ candidates.

\begin{figure}[!htbp]
\centering
\includegraphics[width=0.48\textwidth]{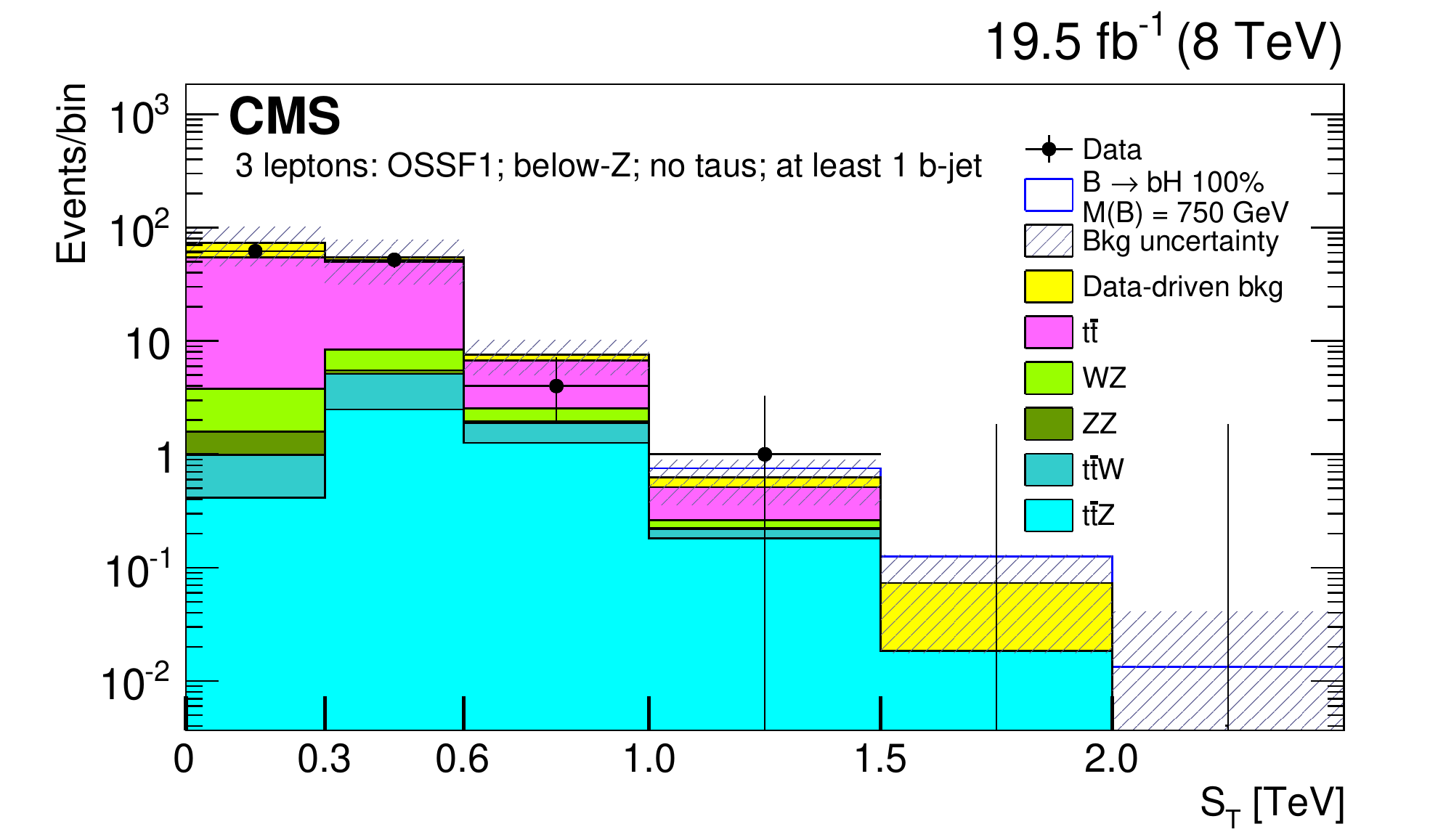} \\
\includegraphics[width=0.48\textwidth]{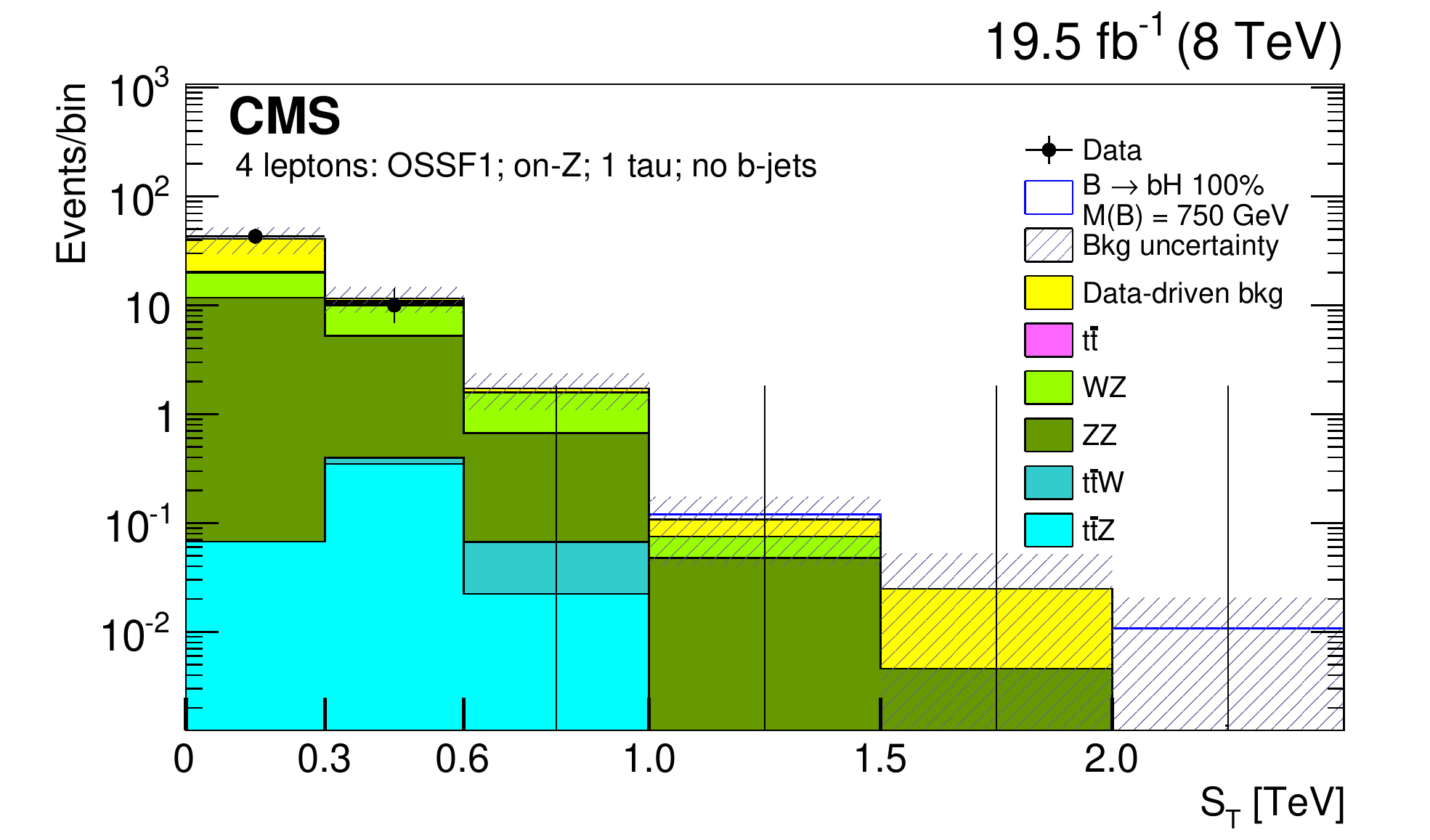}
\caption{Distributions of \ST  for two event categorizations in the multilepton channel: (top) three leptons, no tau leptons, at least one \PQb-tagged jet, and no reconstructed $\Z$ boson candidate; (bottom) four leptons, one tau lepton, no \PQb-tagged jets, and one reconstructed $\Z$ boson candidate. Uncertainties include both statistical and systematic contributions.  The data-driven contribution includes contributions from two genuine leptons and one or more misidentified leptons. The horizontal bars on the data points indicate the bin width. }
\label{fig:multilepST}

\end{figure}

Events are divided into categories based on the reconstructed objects, using the background estimates described above.  These categories include the number of identified leptons, number of identified tau leptons, number of \PQb-tagged jets, and number of $\Z$ bosons reconstructed with an OSSF lepton pair.  Figure~\ref{fig:multilepST} shows the \ST  distribution for two of these event categories.  To further discriminate the \bprime{} quark signal from SM background events, the \ST  distribution is divided into several individual bins: 0--0.3, 0.3--0.6, 0.6--1.0, 1.0--1.5, 1.5--2.0, and $>$2.0\TeV.  These bins are chosen such that the SM backgrounds fall mainly in the lowest two \ST  bins, while the signal events occupy the higher \ST  bins.  Each of these bins is used as the basis for a counting experiment in the final analysis; no further shape discrimination is used within an individual \ST  bin.

\subsection{All-hadronic channel}

The dominant background for the all-hadronic channel comes from SM multijet production. The smaller \ttjets background is obtained from simulation, with corrections to account for differences between data and simulation. The nontop multijet background is obtained from the data. For the multijet background, an $ABCD$ method is used to categorize events into four different categories.

First, events are sorted into categories with \PH-tagged jets or ``anti-\PH-tagged'' jets. Events may contain \PH-tagged jets, as described in Sec.~\ref{sec:evtreco_bHbH}. For events not containing \PH-tagged jets, the criteria of anti-\PH-tagged jets are defined as follows. Pruned CA8 jets are selected such that both the pruned subjets have the \PQb-tagging discriminator variable between 0.244 and 0.679. All other criteria are the same as those for the \PH-tagged jets. This sideband of the \PQb-tagging discriminator variable selects jets enriched in backgrounds with a negligible contribution from signal events.

Next, the pruned mass of the leading  \PH-tagged jet or anti-\PH-tagged jet defines the second selection.  Higgs boson candidates, \ie, \PH-tagged jets with  $90 < \mpruned < 140\GeV$ form the first category of events. This is the signal region, labeled as $B$. Events with $\mpruned < 80\GeV$ form the second category, labeled as $A$. Likewise anti-\PH-tagged events are categorized into the $C$ and $D$ categories depending on whether the leading anti-\PH-tagged jet satisfies $90 < \mpruned < 140\GeV$ or $\mpruned < 80\GeV$, respectively.

Since the event classification variables are uncorrelated, the relation
$N_{B} = N_{A}\,N_{C}/{N_{D}}$ gives the background yields $N_{B}$ in the signal region $B$ based on the yields $N_{A, C, D}$ in the sideband regions $A$, $C$, or $D$, respectively.

Closure tests of the background estimation method were performed separately using simulation and data, with a control sample consisting of events with no \bjets. For the \PQb-jet veto, events with AK5 jets with $\pt > 30\GeV$ and having a \PQb-tagging discriminant value greater than 0.244 are rejected. The \PQb-jet veto criteria applied are stricter than the selection criteria of \bjets used for selecting signal events, thus the control sample is orthogonal to the signal sample. In this case, the region $B$ still contains Higgs candidates and the rest of the sidebands $A$, $C$, and $D$ are as defined above, but without any \bjet.

\begin{figure}[htb]
 \centering
  \includegraphics[width=0.48\textwidth]{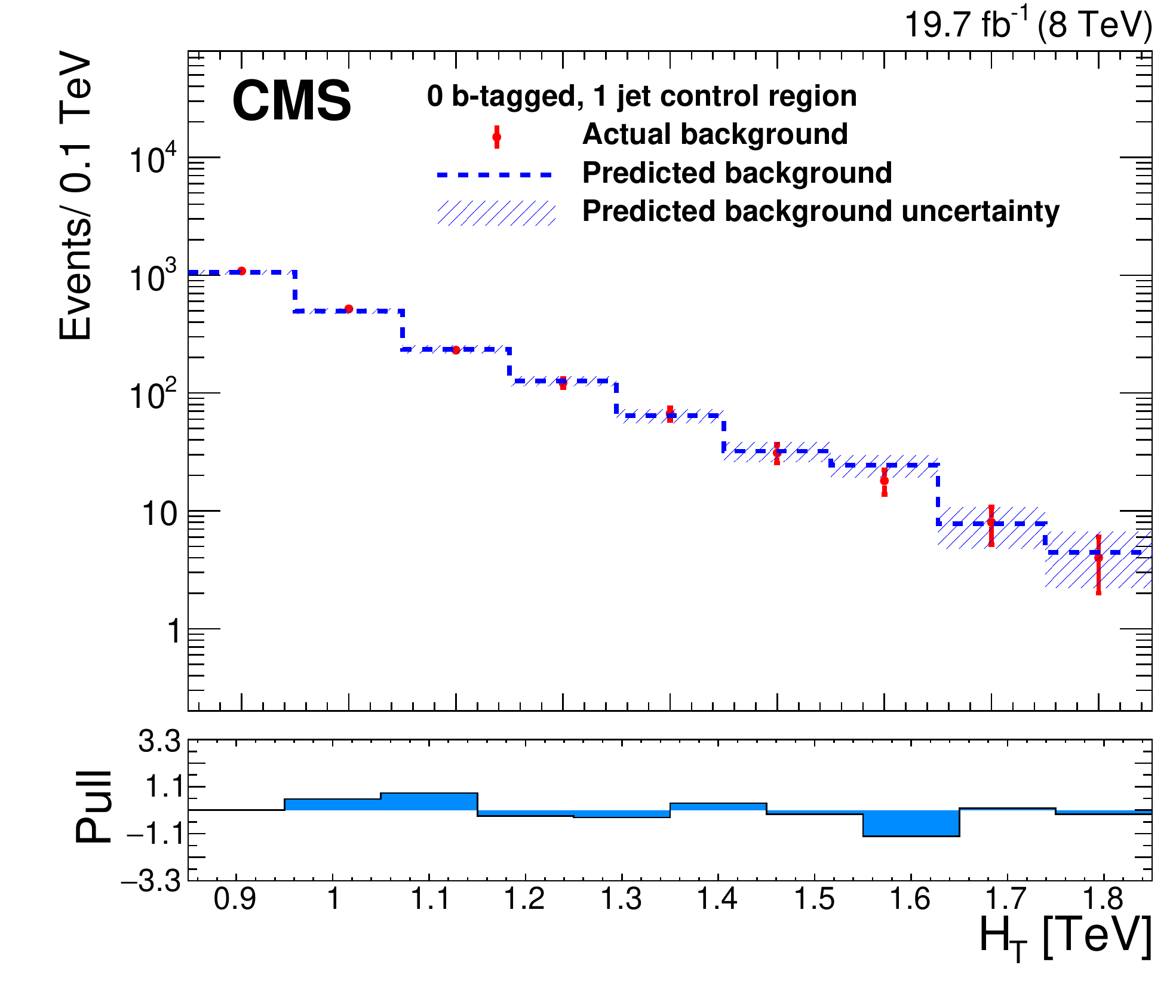}
  \includegraphics[width=0.48\textwidth]{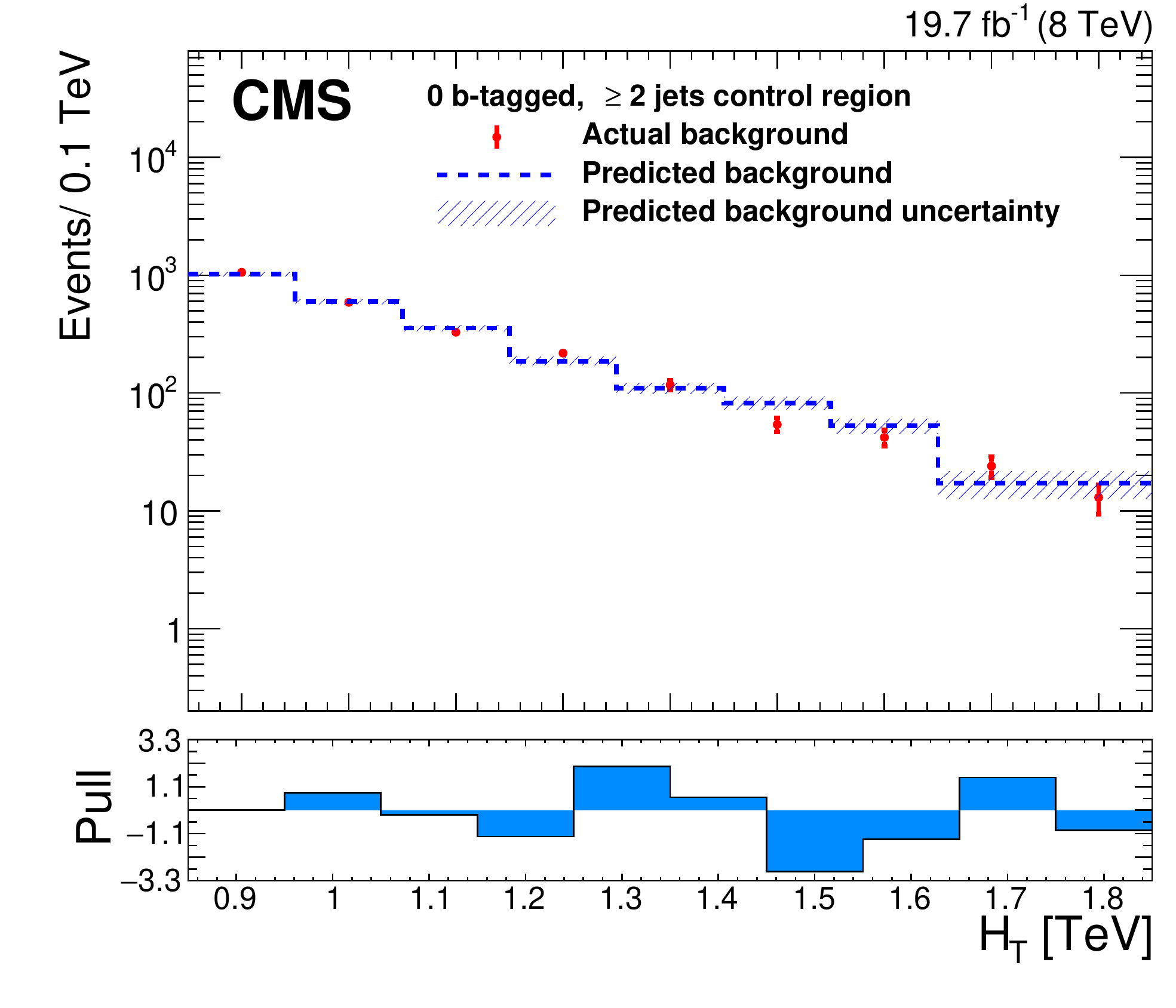}
  \caption{Background closure test in the data control samples of the all-hadronic analysis with no \bjets and exactly one AK5 jet with $\pt > 80\GeV$ (\cmsLeft) and with $\ge$2 AK5 jets with $\pt > 80\GeV$ (\cmsRight). The red data points represent the actual background as derived from data. The blue dashed line represents the predicted background, with the hatched area depicting the corresponding uncertainty. The difference between the observed and predicted background divided by the total uncertainty in the background prediction (pull) is shown for each bin in the lower panels.}
  \label{fig:ABCDClosure_DataControl}

\end{figure}

In the \PQb-jet-vetoed sample, two separate event categories are defined: one with exactly one AK5 jet with $\pt > 80\GeV$, and the other with $\ge$2 AK5 jets with $\pt > 80\GeV$. In both the simulated sample and in the data, the estimated and actual numbers of background events in the signal region $B$ are found to be in agreement. Furthermore, the distributions of \HT for the predicted and the actual background are consistent within measurement uncertainties. The predicted and actual background yields in the data control sample are given in Table~\ref{tab:ABCDClosure_DataControl}, while the agreement in the predicted and actual background \HT distributions is shown in Fig.~\ref{fig:ABCDClosure_DataControl}.

\begin{table*}[!htb]
  \topcaption{Closure test of the background estimation method for the all-hadronic channel, in the data using a control sample with no \bjets. The background is estimated from the sideband regions $A$, $C$, and $D$. The product $N_A\,(N_C/N_D)$ is compared with the actual background observed in region $B$ in the data. The uncertainties correspond to the statistical uncertainties from limited sample sizes. The agreement obtained is within the uncertainties.}
  \label{tab:ABCDClosure_DataControl}
 \centering
  \begin{scotch}{lcc}
 & Yields in 1 jet category & Yields in $\ge$2 jet category \\
   \hline
   Background estimation & $2059 \pm 70$ & $2456 \pm 79$ \\
   True background & 2087 & 2449 \\
\end{scotch}
\end{table*}

\begin{figure}[!htb]
  \centering
    \includegraphics[width=0.48\textwidth]{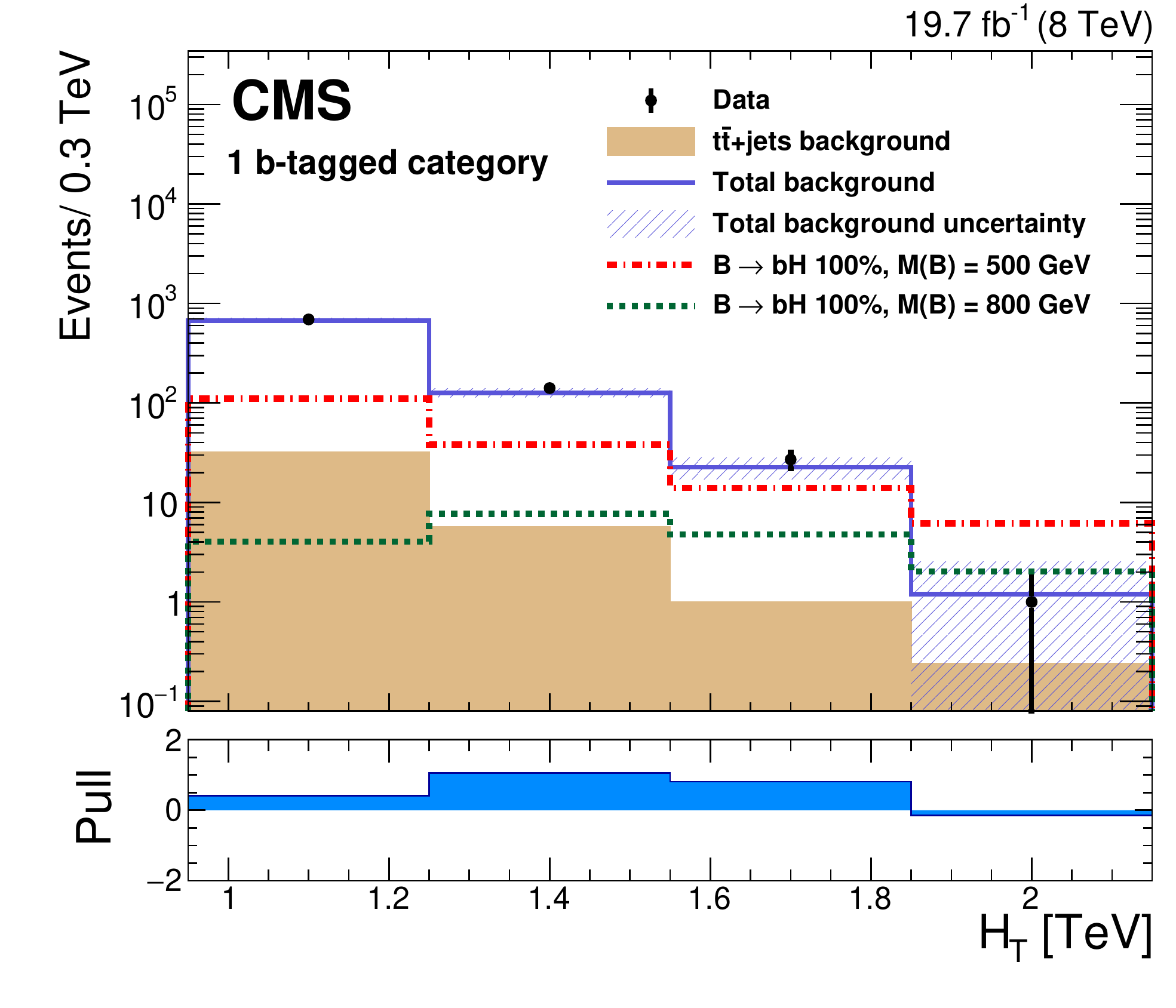}
    \includegraphics[width=0.48\textwidth]{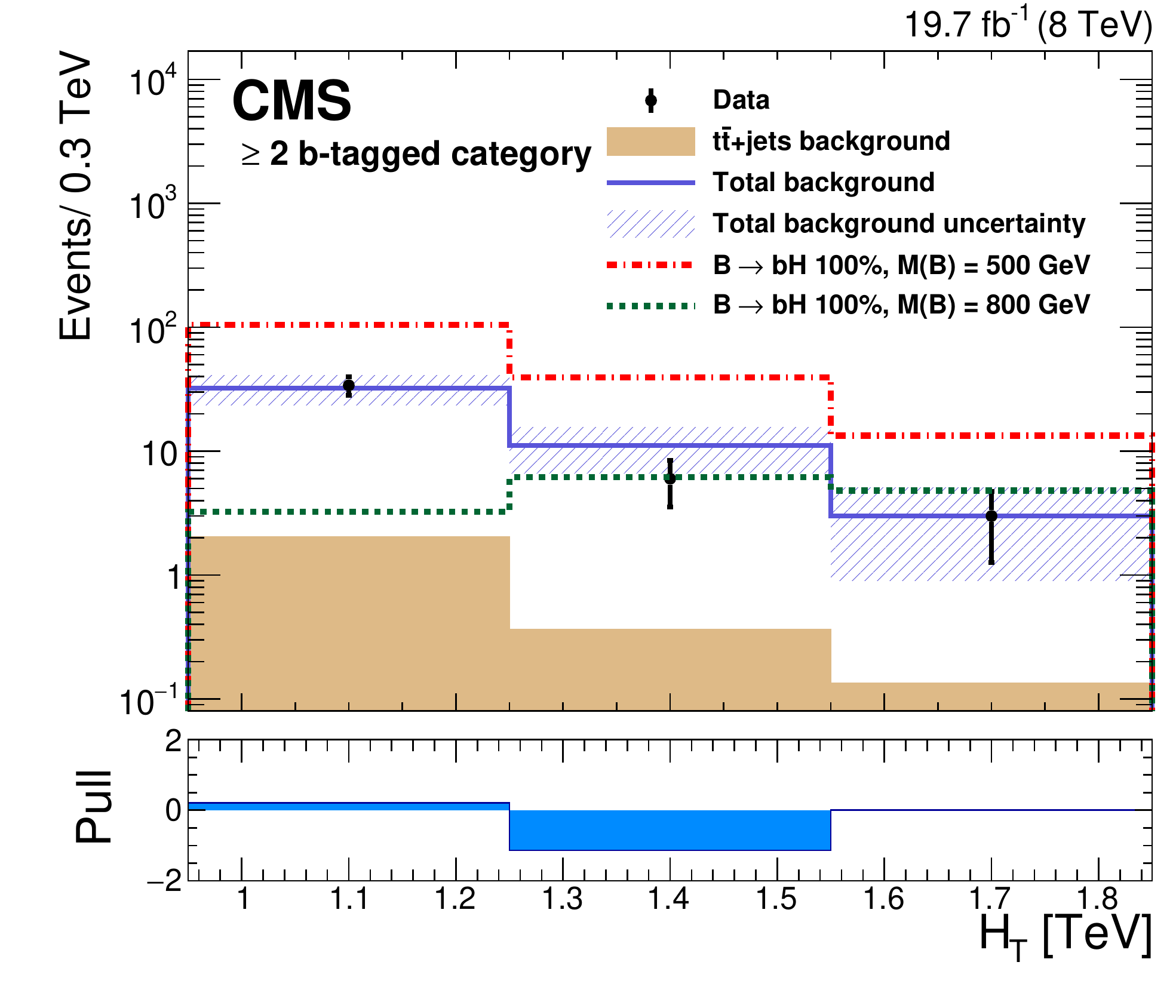}
    \caption{The \HT distributions in the 1 \cPqb{} jet (\cmsLeft) and the  $\ge 2$ \cPqb{} jets event categories (\cmsRight) for the all-hadronic analysis. The blue solid line depicts the estimated background, with the hatched area showing the measured uncertainty in the background. The signal contributions for two \bprime{} quark mass points, 500 and 800\GeV, are overlaid. The bin width is chosen to have a statistically significant number of events in every bin. The difference between the data and the estimated background divided by the total uncertainty in the background (pull) is shown for each bin in the lower panels.}
    \label{fig:HT_BkgFullSyst_Cat1b_Cat2b}
\end{figure}

After closure tests are performed, the $ABCD$ background estimation method is applied to the signal region, and the estimated backgrounds in the 1 \cPqb{} jet, the ${\ge}2$ $\cPqb{}$ jets, and combined categories are shown in Table~\ref{tab:bkg_est}. The background estimated in this way is in agreement with the observed number of events. The \HT distributions in the categories with 1 and ${\ge}2$ $\bjets$ are shown in Fig.~\ref{fig:HT_BkgFullSyst_Cat1b_Cat2b} and show agreement in both shape and normalization with the observed \HT distributions. The uncertainty in the background estimation is propagated from the statistical uncertainties of the samples in the sideband regions and also includes the statistical uncertainties in the control samples. No additional systematic uncertainty has been assigned.  Upper limits on the $\bprime\baprime$ cross section are derived using these distributions.

\begin{table*}[!htb]
  \topcaption{Estimated background and the event yields in the data for the 1 \cPqb{} jet, ${\ge}2$ $\cPqb{}$ jets, and combined event categories in the all-hadronic channel. The uncertainties in the background yields are obtained by propagating the statistical uncertainties from the sideband samples.}
  \label{tab:bkg_est}
 \centering
  \begin{scotch}{lccc}
   & Yields after & Yields in 1 \PQb-tagged & Yields in ${\ge}$2 \PQb-tagged \\
   & full  selection & category & category \\
   \hline \\[-2.2ex]
   Estimated background & $872^{+49}_{-55}$ & $825^{+47}_{-52}$ & $46^{+4}_{-11}$ \\
   Data & 903 & 860 & 43 \\
  \end{scotch}
\end{table*}

\section{Systematic uncertainties}\label{sec:systematics}

Several sources of systematic uncertainty are considered when testing for the presence of \bprime{} quark signal events.  These uncertainties include those associated with detector measurements such as jet and lepton reconstruction and the luminosity determination, as well as theoretical uncertainties in the cross section due to the choice of renormalization and factorization scales.  Finally, there are uncertainties specific to individual analyses, such as those arising from background estimation methods using data.  In this section we detail the sources of systematic uncertainty affecting the various analyses of individual channels of which the results are then combined for the overall result.  Table \ref{corr_nuisances} summarizes these sources.

\begin{table*}
\topcaption{Nuisance parameters applied to the statistical combination.  They are listed separately for each individual channel, and the $\checkmark$ symbol is used if they are applied to that given channel.  If a nuisance parameter is taken as correlated between channels, the \boxCheck{} symbol is shown.  In some cases, several systematic uncertainties are combined into a single nuisance parameter (for example, in the case of combined lepton categories); in such instances, the $\bullet$ symbol is used to denote the presence of a systematic uncertainty combined with others in a distinct nuisance parameter.  The $\sim$ symbol has been used to denote systematic uncertainties that have negligible effects on the analysis results.  The ``Combined systematic uncertainty'' entry represents a contribution composed of other sources listed in the table, applied as a single nuisance parameter during limit extraction.}
\centering
\resizebox{\textwidth}{!}{
\begin{scotch}{lccccc}
 & Lepton+jets  & OS dilepton  & SS dilepton  & Multilepton  & All hadronic\\
\hline
Jet energy scale                & \boxCheck    & \boxCheck    & \boxCheck    & \boxCheck    & \boxCheck   \\
Jet energy resolution           & \boxCheck    & \boxCheck    & \boxCheck    & $\sim$             & \boxCheck   \\
V-tag SF                       & $\checkmark$ &              &              &              & \checkmark  \\
\ttbar matching scale           & $\checkmark$ &              &              &              & $\bullet$     \\
\ttbar renormalization/factorization scales             & $\checkmark$ &              &              &              & $\bullet$     \\
\cPqb{}-tagging SF                       & \boxCheck    & $\bullet$ &              & \boxCheck    & $\bullet$  \\
Light-jet-tagging SF			&               &  $\bullet$ &                 &              & $\bullet$ \\
Integrated luminosity                      & \boxCheck    & \boxCheck    & \boxCheck    & \boxCheck      & \boxCheck   \\
Lepton reconstruction          & $\checkmark$ & $\checkmark$ & $\checkmark$ & $\bullet$      &             \\
\ttbar cross section            & \boxCheck    &              &              & \boxCheck    &             \\
QCD normalization               & $\checkmark$ &              &              &              &             \\
Trigger efficiency              & \checkmark   & $\checkmark$ & $\checkmark$ & $\bullet$      & \checkmark  \\
Pileup uncertainty              &  $\sim$            &  \boxCheck   & \boxCheck    &   $\sim$           & \boxCheck   \\
Background component from data  &              & $\checkmark$ & $\checkmark$ &  $\bullet$     &             \\
PDF uncertainty                 &  $\sim$            &   $\sim$           & $\checkmark$ & $\bullet$      & \checkmark  \\
\ETslash resolution                  &              &              &              & \checkmark   &             \\
Initial-state radiation         &              &              &              & \checkmark   &             \\
Combined systematic uncertainty & $\checkmark$ &              & $\checkmark$ & \checkmark   & \checkmark  \\
\end{scotch}
}
\label{corr_nuisances}
\end{table*}

The jet energy scale and the jet energy resolution uncertainties are taken as fully correlated between each of the individual channels.  These uncertainties are associated with the calibration of the jet energy response in the detector readout.  This calibration procedure, which is dependent on jet \pt and pseudorapidity, leads to an approximately $10\%$ uncertainty in the normalization of event yields.  This uncertainty is applied on an event-by-event basis, resulting in an additional shape effect.  As we observe a difference between the simulated jet energy resolution and the jet energy resolution measured in data, we smear the jet energies in simulation to reflect the energy resolution observed in data.  This procedure introduces a small uncertainty in the shape of jet kinematic properties in simulated events, including a normalization effect of less than 5\%.

In addition to the uncertainty in the measurement of integrated luminosity of 2.6\% \cite{lumi}, several scale factors (SF) are applied to simulated events to reflect the differences with data in reconstruction efficiencies for various objects used in the event selections.  The uncertainties in these SF measurements are applied to the relevant events.   These uncertainties vary for individual channels, but they include SF uncertainties for the electron and muon identification and efficiency values, typically 1\%-2\%, as well as SF uncertainties for the \PQb-tagging algorithms used, at approximately 5\%, and finally specific SF uncertainties for the identification of hadronically decaying high-\pt \PW, \Z, or Higgs bosons, which can be up to 10\% depending on the algorithm used, resulting from the number of events used to measure the appropriate SFs.

For simulated \ttbar events, several specific systematic uncertainties are applied to cover differences in generation parameters.  The renormalization and factorization scales are varied up and down coherently by a factor of 2 to produce a shape uncertainty for the simulated \ttbar events.  This shape template has a normalization effect of roughly 20\%, in addition to the shape component.  The scale used for the parton matching in \PYTHIA is changed to measure an additional systematic uncertainty.  This component is smaller, and is about a 10\% effect.  Finally, an uncertainty of 15\% is applied resulting from the measurement of the \ttbar cross section \cite{Khachatryan:2015oqa}.  This is applied as purely a normalization effect.  These uncertainties only apply to simulated \ttbar events.

We also include systematic uncertainties associated with the choice of the CTEQ6L1 PDF set.  These are estimated by applying weighting factors to vary the elements of the eigenvector used in the PDF simulation.  The weights are combined in quadrature to compute a total systematic uncertainty in the shape and normalization of simulated \ttbar and signal events due to PDF effects.

Finally, there are several uncertainties specific to individual channels.  In some cases, all relevant uncertainties are combined into a single nuisance parameter affecting the normalization, for example for the multilepton channels, which include several counting experiments without shape effects.  Other uncertainties include those for background estimates from data, and are detailed in the corresponding previous sections.

For the statistical combination, we correlate the systematic uncertainties that arise from the same physical effect or phenomenon, such as the jet energy scale, luminosity measurement, or \ttbar cross section.  These correlations allow us to better constrain the uncertainties by using independent information from various channels.  This procedure further improves the sensitivity of the combination.

\section{Combination strategy}\label{sec:combination}

Each of the five channels has a distinct method to discriminate the \bprime quark signal from the expected background contribution.  In the case of the lepton+jets channel, the \ST  distribution is used (Fig.~\ref{fig:ljets_ST}), with different categories corresponding to unique numbers of merged vector bosons reconstructed in the final state.  For the opposite-sign dilepton channel, the \bprime quark candidate mass is reconstructed, and its distribution is used to discriminate the signal (Fig.~\ref{fig:bkgDataDriven_8TeV}).  In the case of the same-sign dilepton and multilepton channels, for each of the various event categorizations, the \ST variable is used for signal discrimination, and each bin of the \ST distribution (Figs.~\ref{fig:ssdil_ST} and \ref{fig:multilepST}) is treated as an independent counting experiment. The results are combined to produce a cross section limit. Finally, the all-hadronic channel uses the \HT distribution separately for single- and double-\PQb-tagged events (Fig.~\ref{fig:HT_BkgFullSyst_Cat1b_Cat2b}).

We combine all signal bins of the five individual analysis channels for the result.  A joint likelihood maximization is performed, simultaneously using the background and signal expectations in each bin, to extract the final results using a Bayesian approach.  We scan over the entire parameter space of the $\bprime$ quark branching fractions in steps of 0.1 for each possible $\bprime$ quark decay mode.

Nuisance parameters are included in the joint likelihood maximization to account for the various systematic uncertainties described above.  For those uncertainties that arise from the same physical or detector effect and are shared between individual channels, the corresponding nuisance parameters are taken to be 100\% correlated in the fitting procedure.  All nuisance parameters describing systematic uncertainties are implemented either with Gaussian priors (for normalization effects) or through template interpolation (for shape changing effects).  The parameter governing the signal normalization is implemented with a uniform prior distribution.

No significant excess above SM expectations is observed.  We set limits on the $\bprime\baprime$ production cross section using the combination of all individual channels to further improve the sensitivity to this process.

\section{Results}\label{sec:results}

Since the vectorlike \bprime quark can decay in three possible topologies ($\PQt\PW$, $\PQb\Z$, and $\PQb\PH$), we scan over the entire possible parameter space of decays, using steps of 10\% in each branching fraction.  This results in 66 combinations of branching fractions, each with its own cross section limits as a function of the $\bprime$ quark mass.  The results are shown in the form of limits on the $\bprime$ quark pair-production cross section and are quoted at 95\% \CLns.

\begin{figure*}[htb]
\centering
\includegraphics[width=0.32\textwidth]{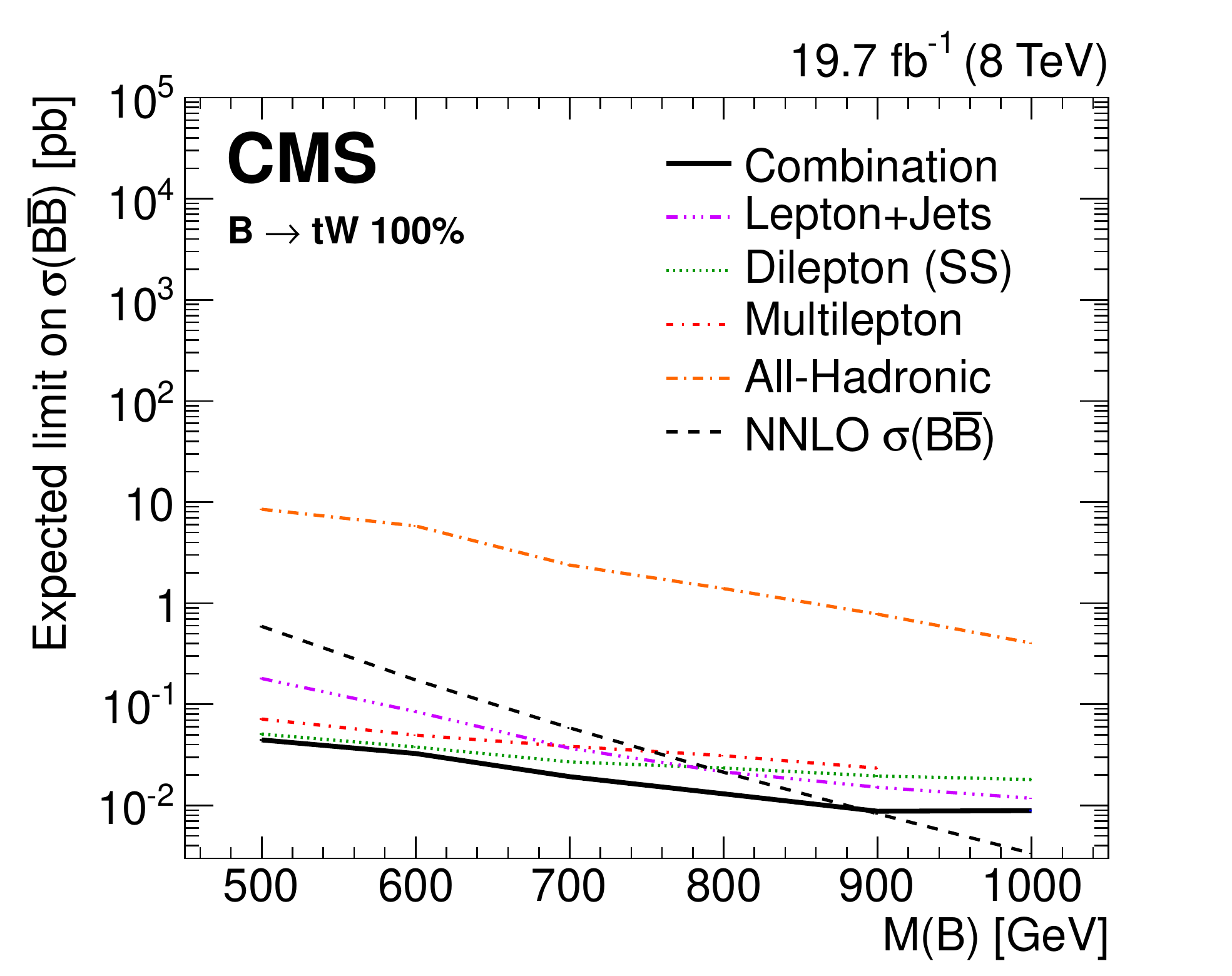}
\includegraphics[width=0.32\textwidth]{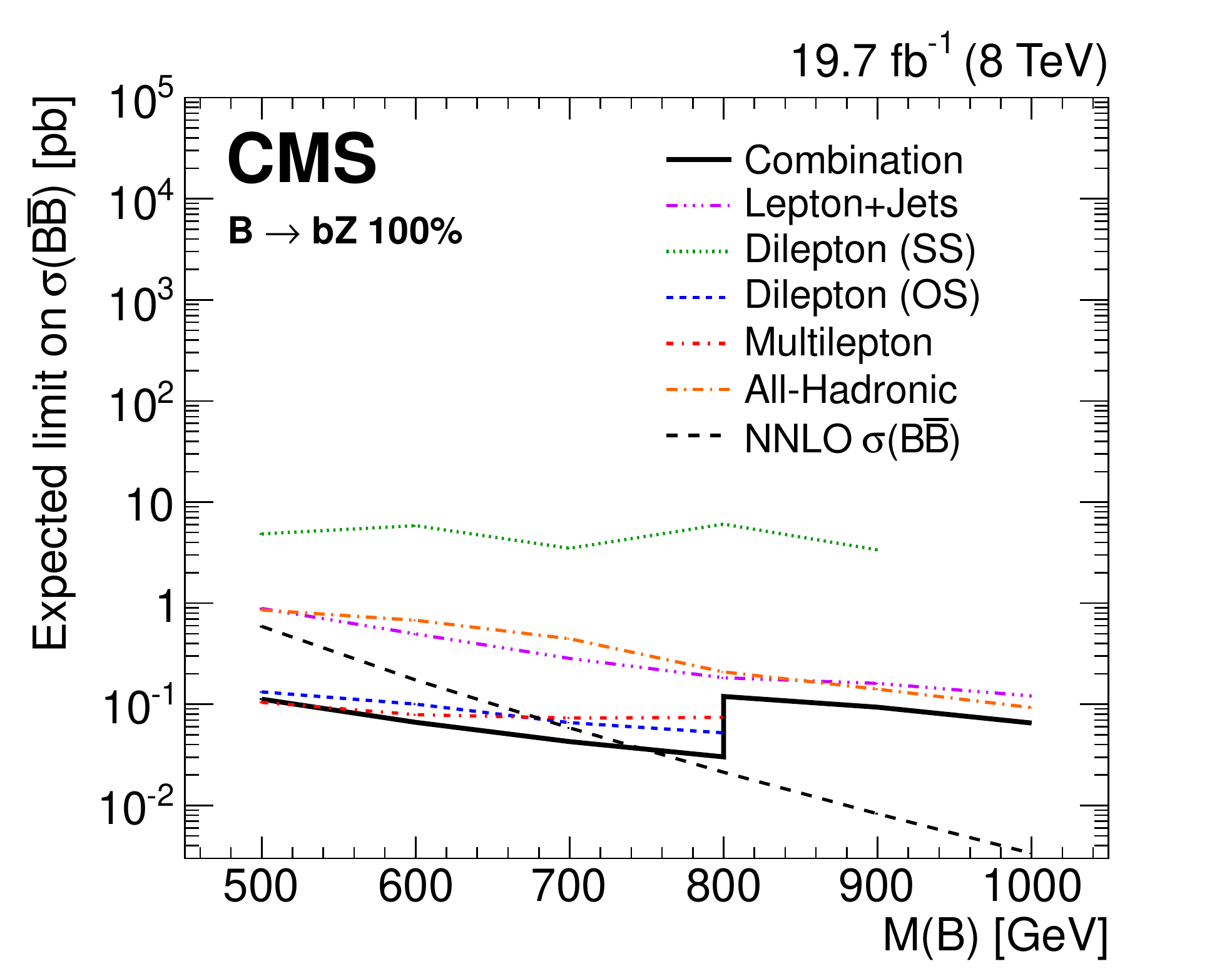}
\includegraphics[width=0.32\textwidth]{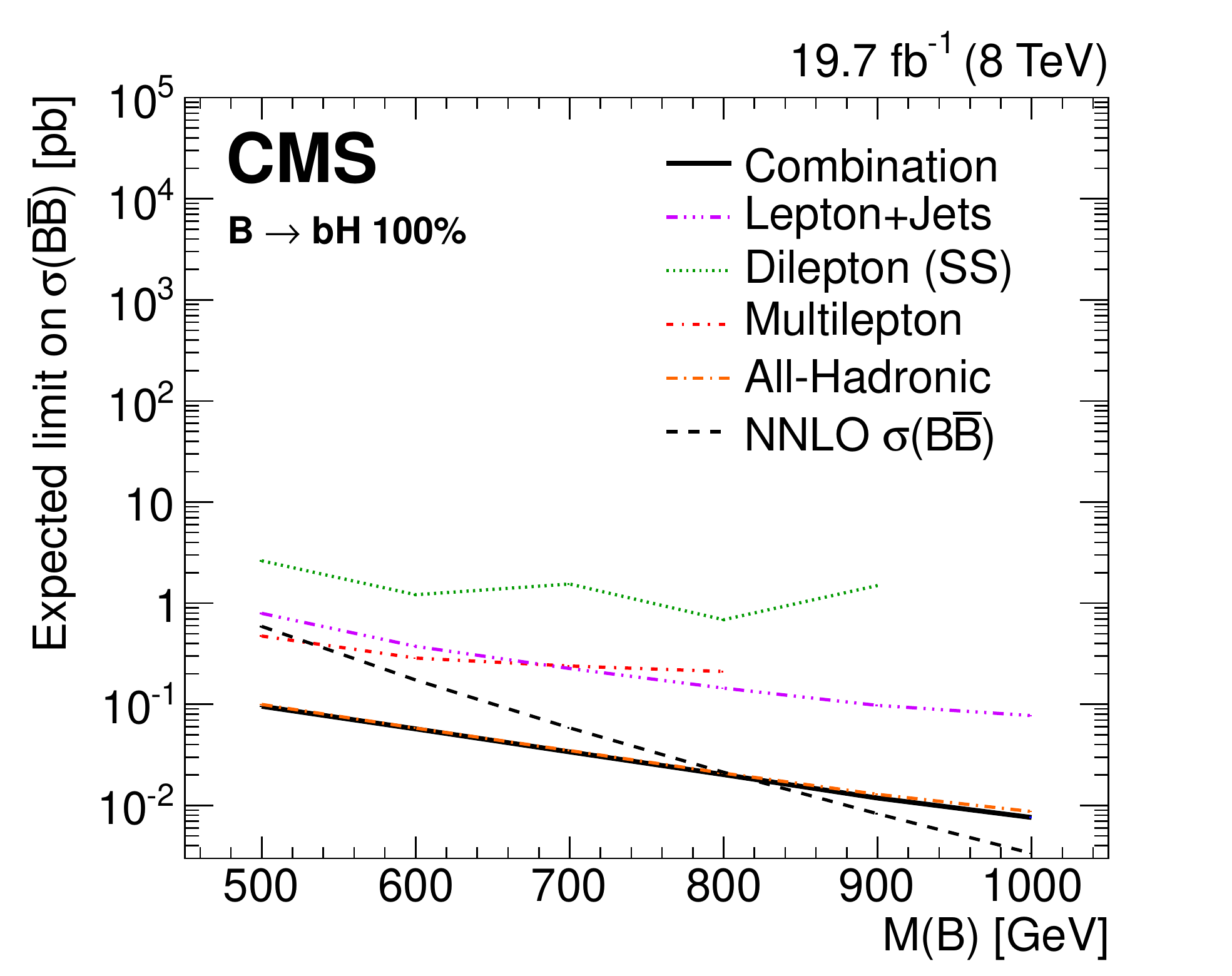}
\caption{Comparison of individual channel limit results with those of the combination, for the expected limits only.  Shown are the limit results, at 95\% \CL, for the three corners of the triangular parameter space, for 100\% branching fractions for $\bprime$ to $\PQt\PW$ (left), $\PQb\Z$ (middle), and $\PQb\PH$ (right).  The same-sign and opposite-sign channels are denoted by ``SS'' and ``OS'', respectively.  The step observed in the $\PQb\Z$ limit curve is due to two analysis channels (multilepton and OS dilepton) that do not contribute to the combination for \bprime quark masses above 800\GeV.}
\label{fig:contributions}
\end{figure*}

The various channels have targeted different final-state topologies, and thus they will contribute to separate regions of the parameter space of decay possibilities.  For example, the lepton+jets channel is sensitive to the $\PQt\PW$ decay mode but is less sensitive to $\PQb\PH$ and $\PQb\Z$ final states.  The opposite-sign dilepton channel is sensitive to the $\PQb\Z$ decay but less so to $\PQb\PH$ or $\PQt\PW$.  The relative contributions of each channel in the case of a 100\% branching fraction for a specific decay mode are shown in Fig.~\ref{fig:contributions}.  In some cases, a channel does not have sensitivity to a certain decay topology and is not included at all in the corresponding result.

\begin{figure*}[htb]
\centering
\includegraphics[width=\cmsFigWidth]{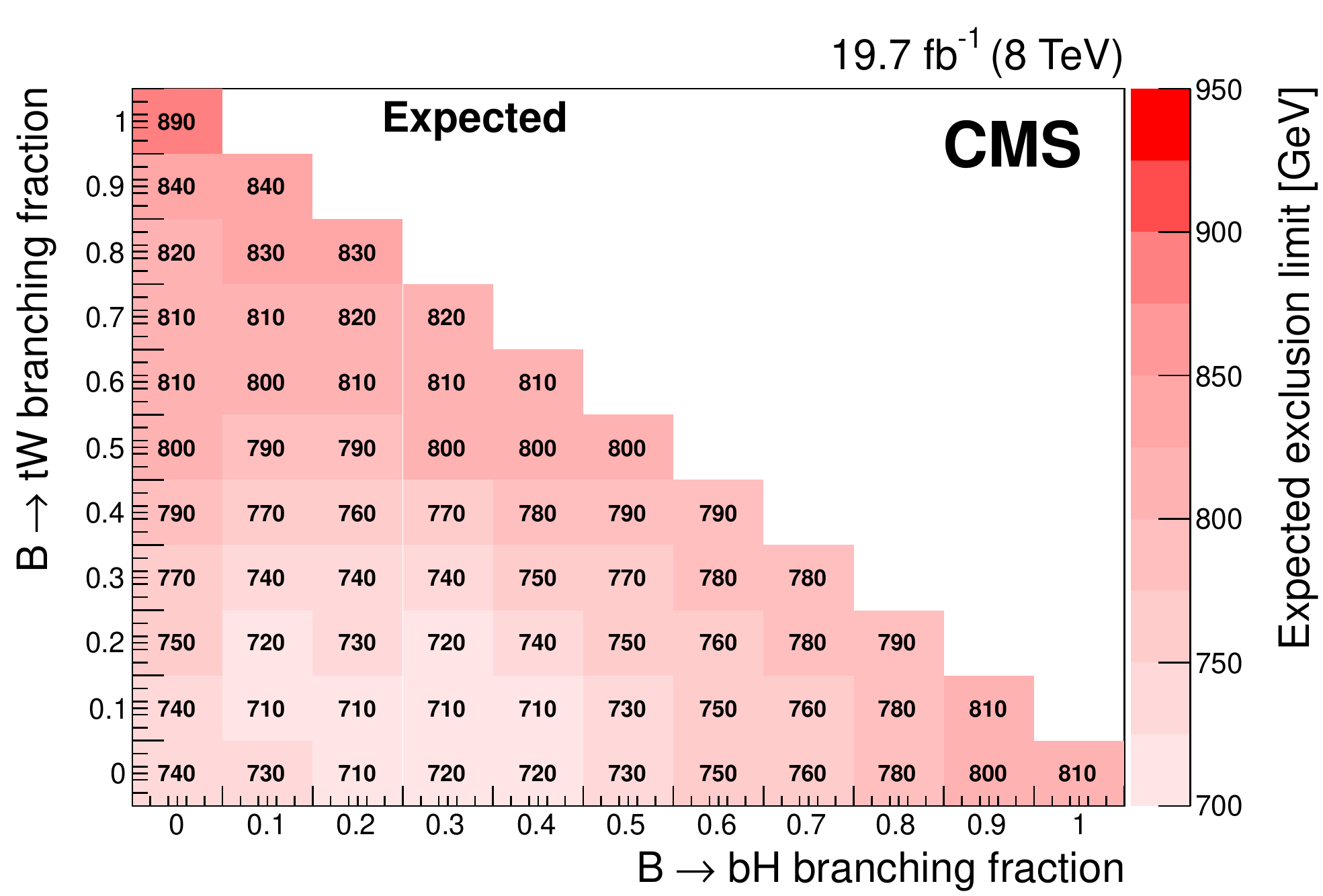}
\includegraphics[width=\cmsFigWidth]{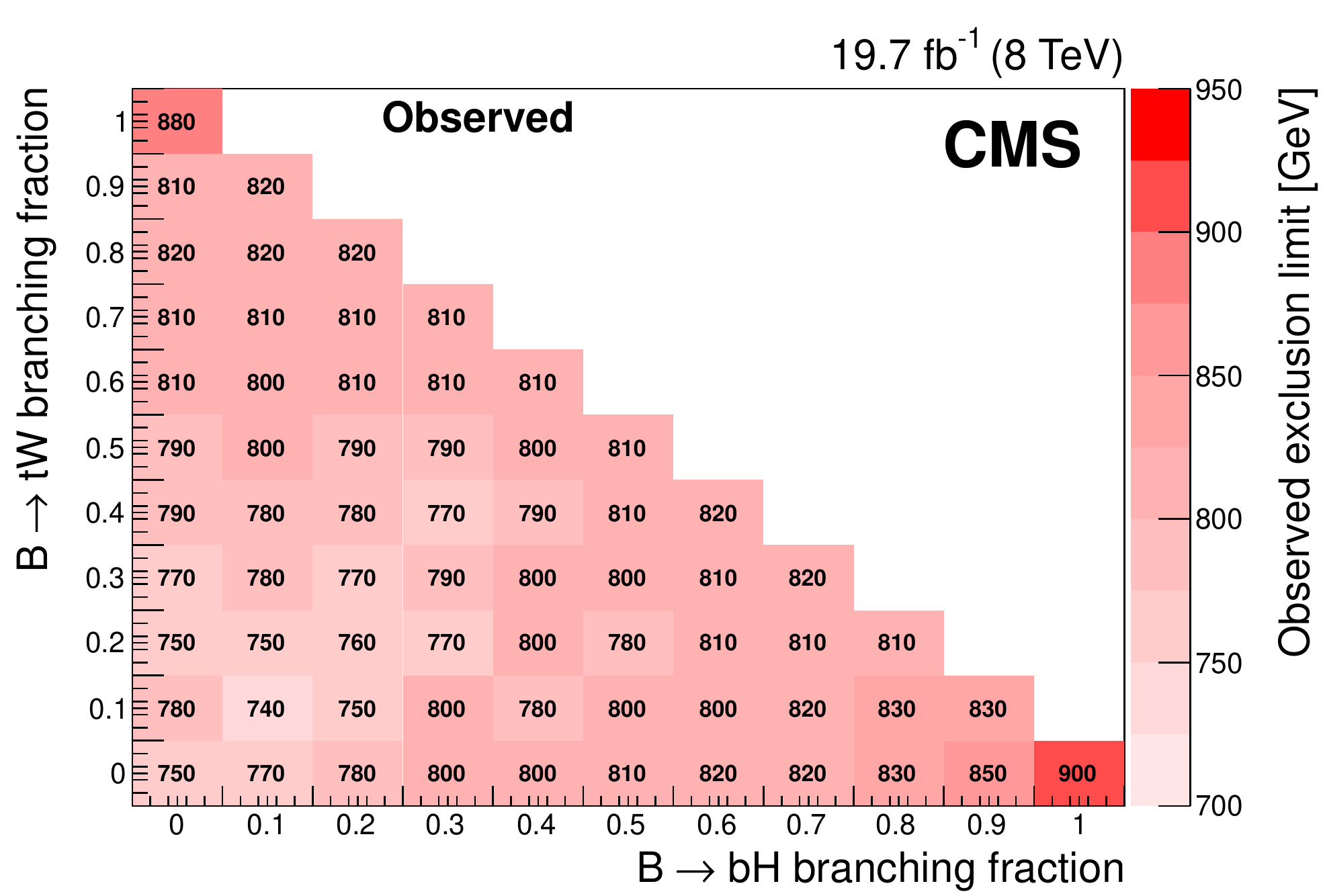}
\caption{Expected (left) and observed (right) limits for each combination of branching fractions to $\PQt\PW$, $\PQb\Z$, and $\PQb\PH$ obtained by the combination of channels.  The color scale represents the mass exclusion limit obtained at each point.  The branching fraction for the \bprime quark decay to $\PQb\Z$ can be obtained through the relation $\mathcal{B}(\PQb\Z) = 1 - \mathcal{B}(\PQt\PW) - \mathcal{B}(\PQb\PH)$.}
\label{fig:limits}
\end{figure*}

The expected and observed exclusion limits for the \bprime quark are determined for each one of the 66 combinations of branching fractions.  To visualize these results, we plot the \bprime quark mass exclusion limits on a triangular parameter space, as shown in Fig.~\ref{fig:limits}.  We also list the cross section limits for different branching fractions and \bprime quark masses in Table \ref{tab:xslimits}. Further details, including event yields for the various channels entering the combination, are available on HepData~\cite{hepdata}.

\begin{table*}
\topcaption{Cross section limits for various combinations of branching fractions and \bprime quark masses.  The expected cross section limits are given in the first row for each branching fraction combination, along with their corresponding uncertainties.  The observed cross section limits are shown in the second row.  All limits are given at 95\% \CL, and are shown in units of pb.}
\centering
\renewcommand*{\arraystretch}{1.2}
\resizebox{\textwidth}{!}{
\begin{scotch}{ccccccccc}
$\mathcal{B}$ & $\mathcal{B}$ & $\mathcal{B}$ & \multicolumn{6}{c}{\bprime quark mass [\GeVns{}]} \\ \cline{4-9}
($\PQt\PW$) & ($\PQb\PH$) & ($\PQb\Z$) & 500 & 600 & 700 & 800 & 900 & 1000 \\
\hline

0.0 & 0.0 & 1.0
& $0.112^{+0.089}_{-0.042}$ & $0.066^{+0.041}_{-0.027}$ & $0.043^{+0.027}_{-0.016}$ & $0.030^{+0.020}_{-0.011}$ & $0.094^{+0.053}_{-0.036}$ & $0.065^{+0.036}_{-0.023}$ \\
& & &  0.090 & 0.037 & 0.036 & 0.029 & 0.096 & 0.083 \\

0.0 & 0.2 & 0.8
& $0.261^{+0.130}_{-0.115}$ & $0.097^{+0.060}_{-0.043}$ & $0.053^{+0.034}_{-0.025}$ & $0.033^{+0.023}_{-0.014}$ & $0.044^{+0.030}_{-0.016}$ & $0.032^{+0.020}_{-0.012}$ \\
& & &  0.215 & 0.046 & 0.038 & 0.023 & 0.044 & 0.033 \\

0.0 & 0.4 & 0.6
& $0.267^{+0.138}_{-0.114}$ & $0.107^{+0.065}_{-0.050}$ & $0.052^{+0.034}_{-0.025}$ & $0.030^{+0.022}_{-0.015}$ & $0.028^{+0.020}_{-0.011}$ & $0.020^{+0.012}_{-0.008}$ \\
& & &  0.210 & 0.065 & 0.037 & 0.020 & 0.027 & 0.021 \\

0.0 & 0.6 & 0.4
& $0.286^{+0.141}_{-0.141}$ & $0.107^{+0.067}_{-0.054}$ & $0.047^{+0.035}_{-0.024}$ & $0.026^{+0.023}_{-0.013}$ & $0.020^{+0.013}_{-0.009}$ & $0.014^{+0.009}_{-0.006}$ \\
& & &  0.219 & 0.058 & 0.031 & 0.016 & 0.019 & 0.014 \\

0.0 & 0.8 & 0.2
& $0.268^{+0.145}_{-0.146}$ & $0.101^{+0.070}_{-0.056}$ & $0.045^{+0.031}_{-0.024}$ & $0.022^{+0.014}_{-0.012}$ & $0.016^{+0.010}_{-0.007}$ & $0.011^{+0.008}_{-0.004}$ \\
& & &  0.131 & 0.061 & 0.030 & 0.016 & 0.014 & 0.010 \\

0.0 & 1.0 & 0.0
& $0.096^{+0.054}_{-0.037}$ & $0.057^{+0.029}_{-0.023}$ & $0.034^{+0.021}_{-0.015}$ & $0.020^{+0.014}_{-0.009}$ & $0.012^{+0.008}_{-0.005}$ & $0.008^{+0.007}_{-0.003}$ \\
& & &  0.066 & 0.035 & 0.021 & 0.014 & 0.008 & 0.006 \\

0.2 & 0.0 & 0.8
& $0.183^{+0.095}_{-0.084}$ & $0.073^{+0.049}_{-0.027}$ & $0.040^{+0.028}_{-0.016}$ & $0.028^{+0.018}_{-0.011}$ & $0.046^{+0.022}_{-0.015}$ & $0.036^{+0.015}_{-0.013}$ \\
& & &  0.140 & 0.041 & 0.033 & 0.029 & 0.058 & 0.046 \\

0.2 & 0.2 & 0.6
& $0.226^{+0.149}_{-0.101}$ & $0.087^{+0.067}_{-0.037}$ & $0.049^{+0.034}_{-0.020}$ & $0.032^{+0.023}_{-0.013}$ & $0.033^{+0.017}_{-0.013}$ & $0.024^{+0.012}_{-0.009}$ \\
& & &  0.099 & 0.057 & 0.035 & 0.028 & 0.045 & 0.031 \\

0.2 & 0.4 & 0.4
& $0.275^{+0.123}_{-0.125}$ & $0.097^{+0.061}_{-0.044}$ & $0.048^{+0.036}_{-0.023}$ & $0.027^{+0.022}_{-0.011}$ & $0.024^{+0.014}_{-0.010}$ & $0.018^{+0.010}_{-0.007}$ \\
& & &  0.131 & 0.049 & 0.032 & 0.021 & 0.027 & 0.022 \\

0.2 & 0.6 & 0.2
& $0.296^{+0.145}_{-0.131}$ & $0.091^{+0.064}_{-0.045}$ & $0.044^{+0.033}_{-0.022}$ & $0.025^{+0.018}_{-0.012}$ & $0.019^{+0.011}_{-0.008}$ & $0.014^{+0.007}_{-0.005}$ \\
& & &  0.162 & 0.047 & 0.031 & 0.019 & 0.021 & 0.015 \\

0.2 & 0.8 & 0.0
& $0.267^{+0.151}_{-0.139}$ & $0.084^{+0.064}_{-0.045}$ & $0.040^{+0.028}_{-0.019}$ & $0.022^{+0.017}_{-0.010}$ & $0.015^{+0.010}_{-0.006}$ & $0.010^{+0.007}_{-0.004}$ \\
& & &  0.107 & 0.052 & 0.023 & 0.018 & 0.015 & 0.012 \\

0.4 & 0.0 & 0.6
& $0.205^{+0.118}_{-0.091}$ & $0.071^{+0.045}_{-0.028}$ & $0.035^{+0.025}_{-0.014}$ & $0.023^{+0.014}_{-0.009}$ & $0.028^{+0.015}_{-0.009}$ & $0.022^{+0.011}_{-0.007}$ \\
& & &  0.152 & 0.042 & 0.029 & 0.023 & 0.040 & 0.030 \\

0.4 & 0.2 & 0.4
& $0.218^{+0.122}_{-0.094}$ & $0.079^{+0.046}_{-0.032}$ & $0.039^{+0.030}_{-0.015}$ & $0.026^{+0.017}_{-0.010}$ & $0.023^{+0.011}_{-0.008}$ & $0.018^{+0.009}_{-0.006}$ \\
& & &  0.111 & 0.047 & 0.027 & 0.025 & 0.032 & 0.026 \\

0.4 & 0.4 & 0.2
& $0.258^{+0.122}_{-0.105}$ & $0.086^{+0.046}_{-0.038}$ & $0.040^{+0.027}_{-0.019}$ & $0.023^{+0.016}_{-0.010}$ & $0.019^{+0.010}_{-0.006}$ & $0.015^{+0.007}_{-0.005}$ \\
& & &  0.142 & 0.040 & 0.028 & 0.023 & 0.026 & 0.020 \\

0.4 & 0.6 & 0.0
& $0.235^{+0.111}_{-0.102}$ & $0.087^{+0.052}_{-0.040}$ & $0.038^{+0.026}_{-0.018}$ & $0.023^{+0.015}_{-0.010}$ & $0.016^{+0.008}_{-0.006}$ & $0.012^{+0.006}_{-0.004}$ \\
& & &  0.139 & 0.043 & 0.028 & 0.015 & 0.020 & 0.015 \\

0.6 & 0.0 & 0.4
& $0.235^{+0.092}_{-0.097}$ & $0.068^{+0.039}_{-0.030}$ & $0.034^{+0.022}_{-0.014}$ & $0.019^{+0.011}_{-0.008}$ & $0.019^{+0.010}_{-0.006}$ & $0.016^{+0.007}_{-0.005}$ \\
& & &  0.135 & 0.051 & 0.025 & 0.018 & 0.029 & 0.021 \\

0.6 & 0.2 & 0.2
& $0.241^{+0.097}_{-0.095}$ & $0.067^{+0.045}_{-0.028}$ & $0.033^{+0.024}_{-0.015}$ & $0.019^{+0.013}_{-0.007}$ & $0.017^{+0.008}_{-0.006}$ & $0.014^{+0.006}_{-0.005}$ \\
& & &  0.163 & 0.035 & 0.022 & 0.015 & 0.024 & 0.020 \\

0.6 & 0.4 & 0.0
& $0.176^{+0.111}_{-0.079}$ & $0.064^{+0.051}_{-0.025}$ & $0.034^{+0.026}_{-0.015}$ & $0.019^{+0.012}_{-0.008}$ & $0.015^{+0.008}_{-0.005}$ & $0.012^{+0.006}_{-0.004}$ \\
& & &  0.113 & 0.033 & 0.023 & 0.016 & 0.021 & 0.015 \\

0.8 & 0.0 & 0.2
& $0.209^{+0.091}_{-0.095}$ & $0.057^{+0.030}_{-0.022}$ & $0.027^{+0.019}_{-0.011}$ & $0.016^{+0.010}_{-0.007}$ & $0.014^{+0.006}_{-0.005}$ & $0.012^{+0.006}_{-0.004}$ \\
& & &  0.138 & 0.036 & 0.024 & 0.015 & 0.021 & 0.017 \\

0.8 & 0.2 & 0.0
& $0.119^{+0.072}_{-0.049}$ & $0.050^{+0.035}_{-0.021}$ & $0.027^{+0.018}_{-0.011}$ & $0.017^{+0.011}_{-0.006}$ & $0.013^{+0.006}_{-0.004}$ & $0.011^{+0.005}_{-0.003}$ \\
& & &  0.064 & 0.029 & 0.019 & 0.016 & 0.019 & 0.015 \\

1.0 & 0.0 & 0.0
& $0.044^{+0.021}_{-0.015}$ & $0.033^{+0.022}_{-0.012}$ & $0.019^{+0.013}_{-0.007}$ & $0.013^{+0.008}_{-0.005}$ & $0.009^{+0.005}_{-0.003}$ & $0.009^{+0.004}_{-0.003}$ \\
& & &  0.033 & 0.023 & 0.016 & 0.013 & 0.009 & 0.013 \\

\end{scotch}
}

\label{tab:xslimits}
\end{table*}

\begin{figure*}[htb]
\centering
\includegraphics[width=0.32\textwidth]{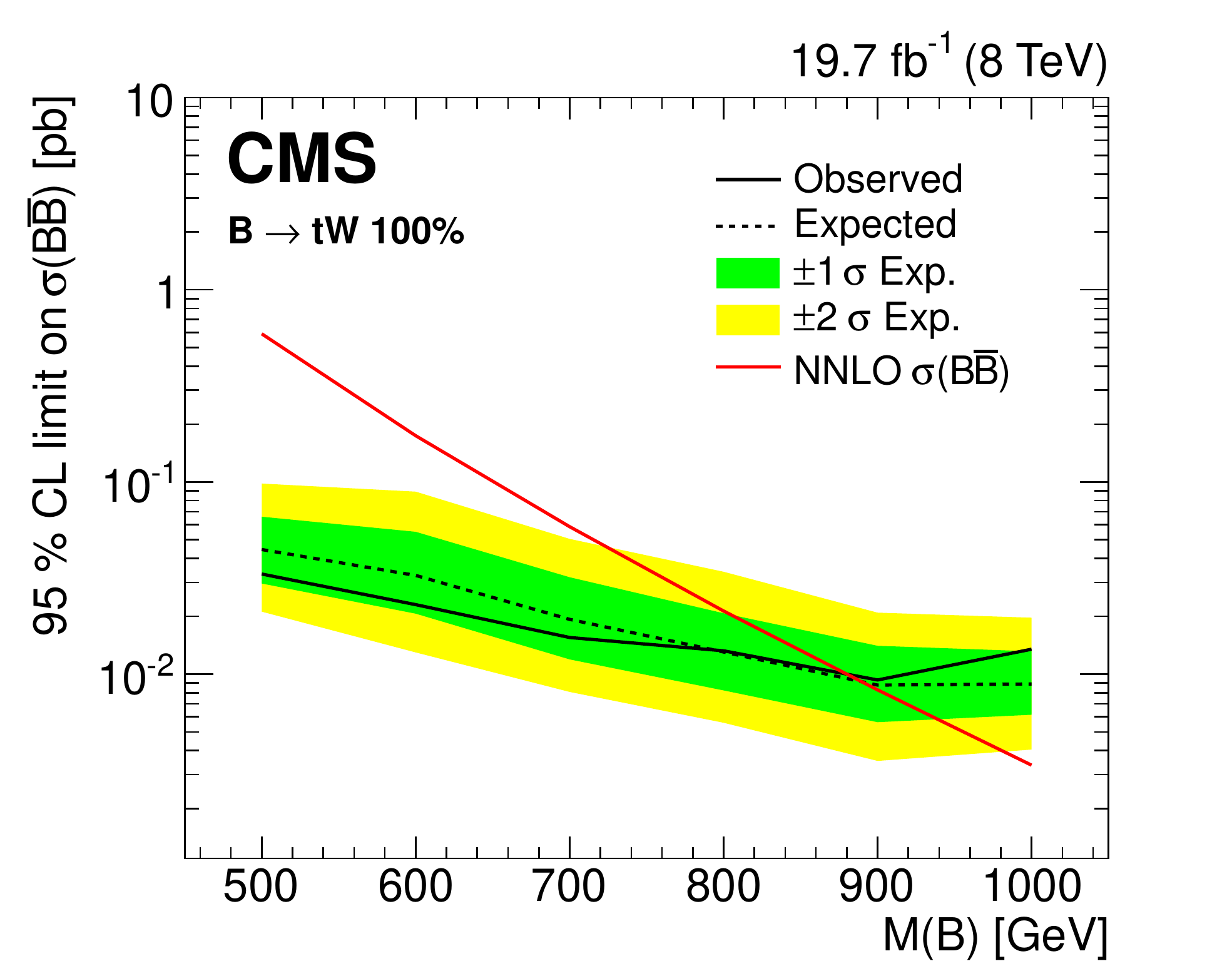}
\includegraphics[width=0.32\textwidth]{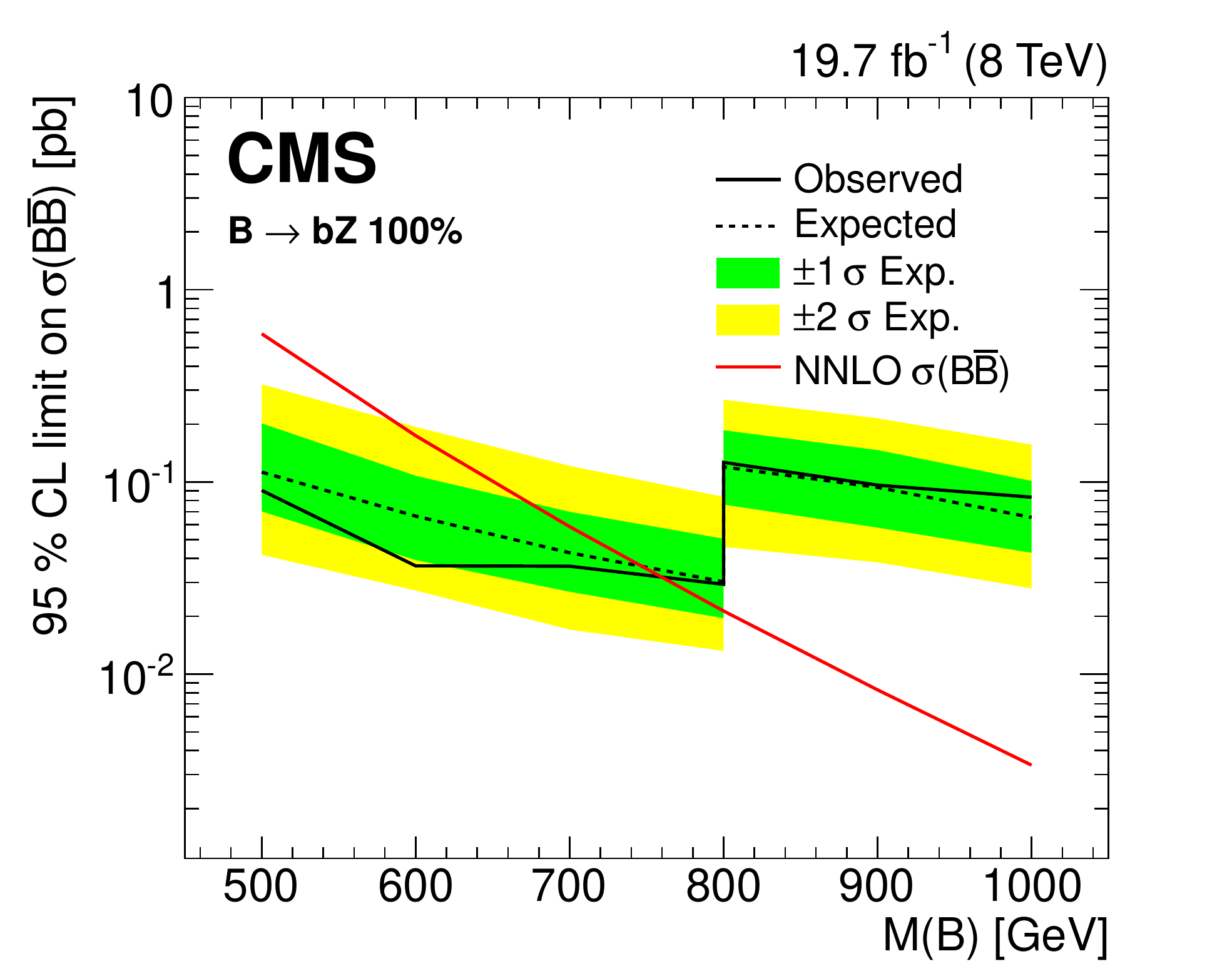}
\includegraphics[width=0.32\textwidth]{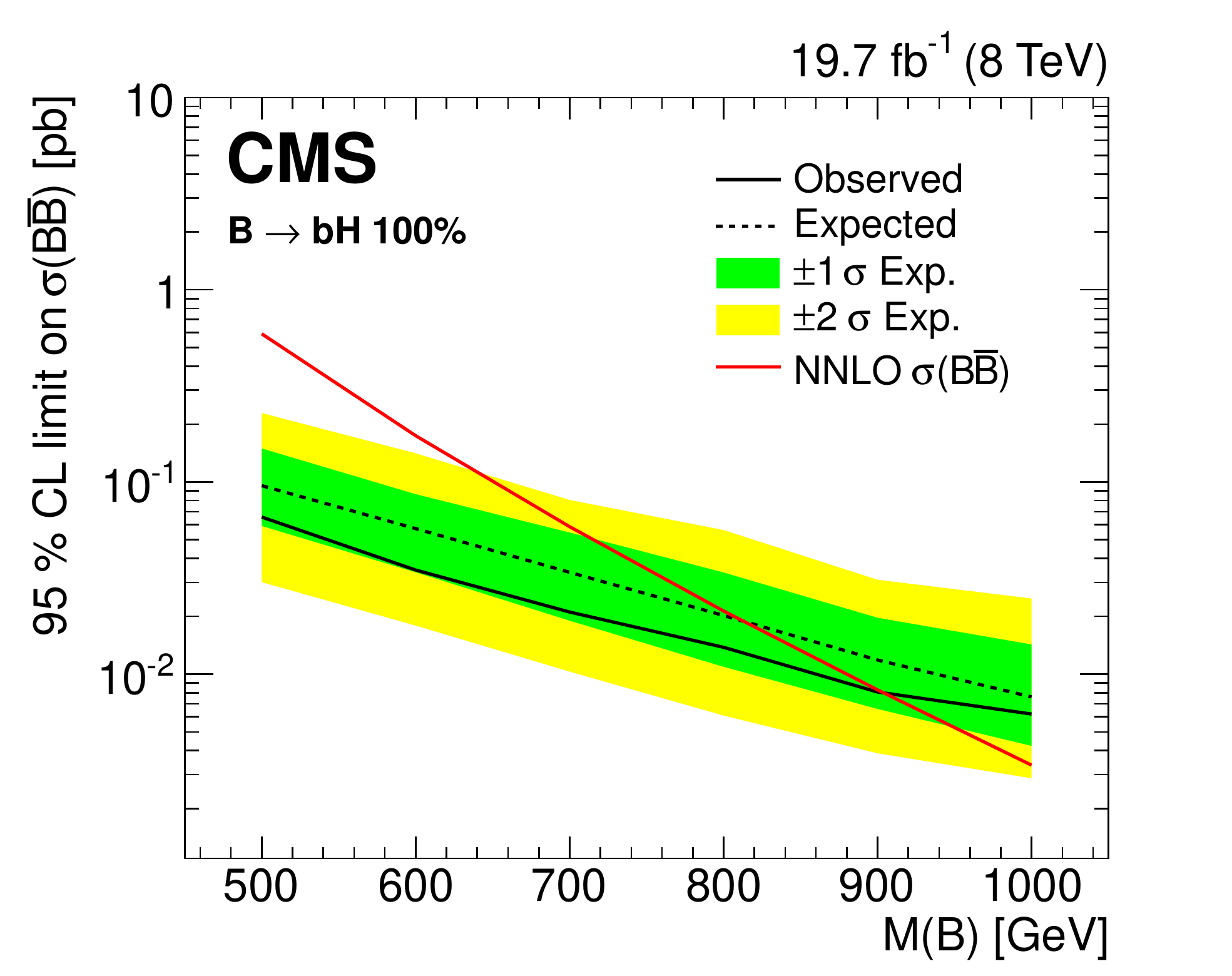}
\caption{Observed and expected cross section limit results as a function of \bprime mass, for the combination of all channels.  The limit results are shown for exclusive branching fractions of \bprime to $\PQt\PW$ (left), $\PQb\Z$ (middle), and $\PQb\PH$ (right).  The step observed in the $\PQb\Z$ limit curve is due to two analysis channels (multilepton and OS dilepton) that do not contribute to the combination for \bprime quark masses above 800\GeV.}
\label{fig:oneDlimits}

\end{figure*}

The expected and observed limits agree within the uncertainties.  For the branching fraction \bprimetotW of 100\%, we expect to exclude $M(\bprime) < \TWEXPLIMIT$\GeV and observe an exclusion of $M(\bprime) < \TWOBSLIMIT$\GeV.  This is the combination with the best sensitivity to \bprime pair production.  The remaining results are summarized in Table \ref{tab:limits}.  Finally, the cross section limits as a function of \bprime{} mass are shown graphically in Fig.~\ref{fig:oneDlimits} for the exclusive decay modes to $\PQt\PW$, $\PQb\Z$, and $\PQb\PH$. The multilepton and OS dilepton channels do not contribute to the combination for \bprime{} quark masses above 800~\GeV. This restriction does not affect the mass exclusion limits obtained from the combined result.

\begin{table}[htb]
\topcaption{Expected and observed mass exclusion limits for the combined result, quoted at 95\% \CL, for the topologies where the \bprime quark decays exclusively in one of the possible modes.}
\centering
\begin{scotch}{ccc}
 & \multicolumn{2}{c}{95\% \CL $M(\bprime)$ exclusion limit [\GeVns{}]} \\ \cline{2-3}
& Expected & Observed \\[\cmsTabSkip]
$\bprime\to\PQt\PW$ & \TWEXPLIMIT & \TWOBSLIMIT \\
$\bprime\to\PQb\PH$ & \BHEXPLIMIT & \BHOBSLIMIT \\
$\bprime\to\PQb\Z$ & \BZEXPLIMIT & \BZOBSLIMIT \\
\end{scotch}
\label{tab:limits}

\end{table}
\section{Summary}
\label{sec:conclusion}

A search for pair production of the \bprime quark with vectorlike couplings to \PW, \Z, and Higgs bosons has been performed, using data recorded by the CMS experiment from proton-proton collisions at a center-of-mass energy of 8\TeV at the CERN LHC in 2012.  This hypothesized particle could decay in one of three ways: to $\PQt\PW$, $\PQb\Z$, or $\PQb\PH$.  The search is performed using five distinct topologies to maintain sensitivity to each of these decay modes.  The topologies included in this search are the lepton+jets final state, both the opposite-sign and same-sign lepton pair final states, the three or more leptons final state, and finally the all-hadronic final state targeting Higgs boson decays to pairs of bottom quarks.

No evidence for the production of \bprime quarks in any topology is found, and limits are set on the $\bprime$ quark-antiquark pair-production cross section.  A scan over possible combinations of the branching fractions to $\PQt\PW$, $\PQb\Z$, and $\PQb\PH$ is performed.  For a $\bprime$ quark decaying with a branching fraction of 100\% to $\PQb\PH$, $\bprime$ quarks with masses up to \BHOBSLIMIT{}\GeV are excluded, at 95\% confidence level.  This branching fraction corresponds to the highest excluded $\bprime$ quark mass in this scan.  Observed exclusion limit results from a scan of all possible branching fractions range from a minimum of 740\GeV to \BHOBSLIMIT{}\GeV.  The combination of these results provides the most stringent exclusion limit to date for the existence of a vectorlike $\bprime$ quark.

\clearpage
\begin{acknowledgments}
\hyphenation{Bundes-ministerium Forschungs-gemeinschaft Forschungs-zentren} We congratulate our colleagues in the CERN accelerator departments for the excellent performance of the LHC and thank the technical and administrative staffs at CERN and at other CMS institutes for their contributions to the success of the CMS effort. In addition, we gratefully acknowledge the computing centers and personnel of the Worldwide LHC Computing Grid for delivering so effectively the computing infrastructure essential to our analyses. Finally, we acknowledge the enduring support for the construction and operation of the LHC and the CMS detector provided by the following funding agencies: the Austrian Federal Ministry of Science, Research and Economy and the Austrian Science Fund; the Belgian Fonds de la Recherche Scientifique, and Fonds voor Wetenschappelijk Onderzoek; the Brazilian Funding Agencies (CNPq, CAPES, FAPERJ, and FAPESP); the Bulgarian Ministry of Education and Science; CERN; the Chinese Academy of Sciences, Ministry of Science and Technology, and National Natural Science Foundation of China; the Colombian Funding Agency (COLCIENCIAS); the Croatian Ministry of Science, Education and Sport, and the Croatian Science Foundation; the Research Promotion Foundation, Cyprus; the Ministry of Education and Research, Estonian Research Council via IUT23-4 and IUT23-6 and European Regional Development Fund, Estonia; the Academy of Finland, Finnish Ministry of Education and Culture, and Helsinki Institute of Physics; the Institut National de Physique Nucl\'eaire et de Physique des Particules~/~CNRS, and Commissariat \`a l'\'Energie Atomique et aux \'Energies Alternatives~/~CEA, France; the Bundesministerium f\"ur Bildung und Forschung, Deutsche Forschungsgemeinschaft, and Helmholtz-Gemeinschaft Deutscher Forschungszentren, Germany; the General Secretariat for Research and Technology, Greece; the National Scientific Research Foundation, and National Innovation Office, Hungary; the Department of Atomic Energy and the Department of Science and Technology, India; the Institute for Studies in Theoretical Physics and Mathematics, Iran; the Science Foundation, Ireland; the Istituto Nazionale di Fisica Nucleare, Italy; the Ministry of Science, ICT and Future Planning, and National Research Foundation (NRF), Republic of Korea; the Lithuanian Academy of Sciences; the Ministry of Education, and University of Malaya (Malaysia); the Mexican Funding Agencies (CINVESTAV, CONACYT, SEP, and UASLP-FAI); the Ministry of Business, Innovation and Employment, New Zealand; the Pakistan Atomic Energy Commission; the Ministry of Science and Higher Education and the National Science Centre, Poland; the Funda\c{c}\~ao para a Ci\^encia e a Tecnologia, Portugal; JINR, Dubna; the Ministry of Education and Science of the Russian Federation, the Federal Agency of Atomic Energy of the Russian Federation, Russian Academy of Sciences, and the Russian Foundation for Basic Research; the Ministry of Education, Science and Technological Development of Serbia; the Secretar\'{\i}a de Estado de Investigaci\'on, Desarrollo e Innovaci\'on and Programa Consolider-Ingenio 2010, Spain; the Swiss Funding Agencies (ETH Board, ETH Zurich, PSI, SNF, UniZH, Canton Zurich, and SER); the Ministry of Science and Technology, Taipei; the Thailand Center of Excellence in Physics, the Institute for the Promotion of Teaching Science and Technology of Thailand, Special Task Force for Activating Research and the National Science and Technology Development Agency of Thailand; the Scientific and Technical Research Council of Turkey, and Turkish Atomic Energy Authority; the National Academy of Sciences of Ukraine, and State Fund for Fundamental Researches, Ukraine; the Science and Technology Facilities Council, UK; the US Department of Energy, and the US National Science Foundation.

Individuals have received support from the Marie-Curie program and the European Research Council and EPLANET (European Union); the Leventis Foundation; the A. P. Sloan Foundation; the Alexander von Humboldt Foundation; the Belgian Federal Science Policy Office; the Fonds pour la Formation \`a la Recherche dans l'Industrie et dans l'Agriculture (FRIA-Belgium); the Agentschap voor Innovatie door Wetenschap en Technologie (IWT-Belgium); the Ministry of Education, Youth and Sports (MEYS) of the Czech Republic; the Council of Science and Industrial Research, India; the HOMING PLUS program of the Foundation for Polish Science, cofinanced from European Union, Regional Development Fund; the Compagnia di San Paolo (Torino); the Consorzio per la Fisica (Trieste); MIUR project 20108T4XTM (Italy); the Thalis and Aristeia programs cofinanced by EU-ESF and the Greek NSRF; the National Priorities Research Program by Qatar National Research Fund; the Rachadapisek Sompot Fund for Postdoctoral Fellowship, Chulalongkorn University (Thailand); and the Welch Foundation.
\end{acknowledgments}

\bibliography{auto_generated}
\cleardoublepage \appendix\section{The CMS Collaboration \label{app:collab}}\begin{sloppypar}\hyphenpenalty=5000\widowpenalty=500\clubpenalty=5000\textbf{Yerevan Physics Institute,  Yerevan,  Armenia}\\*[0pt]
V.~Khachatryan, A.M.~Sirunyan, A.~Tumasyan
\vskip\cmsinstskip
\textbf{Institut f\"{u}r Hochenergiephysik der OeAW,  Wien,  Austria}\\*[0pt]
W.~Adam, E.~Asilar, T.~Bergauer, J.~Brandstetter, E.~Brondolin, M.~Dragicevic, J.~Er\"{o}, M.~Flechl, M.~Friedl, R.~Fr\"{u}hwirth\cmsAuthorMark{1}, V.M.~Ghete, C.~Hartl, N.~H\"{o}rmann, J.~Hrubec, M.~Jeitler\cmsAuthorMark{1}, V.~Kn\"{u}nz, A.~K\"{o}nig, M.~Krammer\cmsAuthorMark{1}, I.~Kr\"{a}tschmer, D.~Liko, T.~Matsushita, I.~Mikulec, D.~Rabady\cmsAuthorMark{2}, B.~Rahbaran, H.~Rohringer, J.~Schieck\cmsAuthorMark{1}, R.~Sch\"{o}fbeck, J.~Strauss, W.~Treberer-Treberspurg, W.~Waltenberger, C.-E.~Wulz\cmsAuthorMark{1}
% ...
\vskip\cmsinstskip
% ...
1:~~Also at Vienna University of Technology, Vienna, Austria\\
2:~~Also at CERN, European Organization for Nuclear Research, Geneva, Switzerland\\
% ...

\end{sloppypar}
\end{document}